\documentclass[showpacs,preprintnumbers,amsmath,amssymb]{article}

\usepackage[english]{babel}
\usepackage{graphicx}
\usepackage{epsfig}
\usepackage{color}
\usepackage{latexsym}
\usepackage{bm}
\usepackage{amsfonts}
\makeatletter
\def\cont{\mathbin{\dimen0=\ht\strutbox \dimen1=\ht\strutbox \divide\dimen0 by 2
\divide\dimen1 by 4
  \hbox{\vbox{\hrule width\dimen1}\hskip-0.4pt\vrule   height\dimen0}}\,}

\newcommand{\Leftarrowfill}[0]{$\m@th  \mathord\Leftarrow  \mkern-6mu
\cleaders\hbox{$\mkern-2mu \mathord= \mkern-2mu$}\hfill
\mkern -6mu \mathord=$}

\newcommand{\StemPullBack}[2]{
  \vtop{\mathsurround=0pt
  \ialign{##\crcr$\textstyle{#1}\strut$\crcr
    \noalign{\kern-0.4ex\nointerlineskip}{\tiny#2}\crcr}}}

\newcommand{\IndxPullBack}[2]{
  \vtop{\mathsurround=0pt
  \ialign{##\crcr\hfil$\scriptstyle{#1}$\hfil\crcr
    \noalign{\kern+0.4ex\nointerlineskip}{\tiny#2}\crcr}}}

\newcommand{\Real}{\mathbb{R}} 

\newcommand{\cN}{{\cal N}}

\newcommand{\ag}{\alpha} 
\newcommand{\bg}{\beta} 
\newcommand{\Dg}{\Delta}
\newcommand{\dg}{\delta} 
\newcommand{\cg}{\gamma} 
\newcommand{\Cg}{\Gamma}

\newcommand{\lam}{\lambda}

\newcommand{\veg}{\varepsilon} 

\newcommand{\Sg}{\Sigma} 
\newcommand{\sg}{\sigma} 

\newcommand{\di}{\partial} 
\newcommand{\be}{\begin{equation}} 
\newcommand{\ee}{\end{equation}} 
\newcommand{\bearr}{\begin{eqnarray}}
\newcommand{\eearr}{\end{eqnarray}}

\newcommand{\QED}{\rule{1.5mm}{3mm}} 
\newcommand{\Cyc}[1]{
  \vtop{\mathsurround=0pt
  \ialign{##\crcr$\textstyle{\rm Cyc}\strut$\crcr
    \noalign{\kern-0.4ex\nointerlineskip}{\tiny#1}\crcr}}\ }

\begin{document}

\title{The symplectic 2-form for gravity in terms of free null initial data}
\author{Michael P. Reisenberger\\
Instituto de F\a'{\i}sica, Facultad de Ciencias,\\
        Universidad de la Rep\a'ublica Oriental del Uruguay,\\
        Igu\a'a 4225, esq. Mataojo, Montevideo, Uruguay}
\date{\today}
\maketitle

\begin{abstract}
A hypersurface $\cN$ formed of two null sheets, or "light fronts", swept out by 
the future null normal geodesics emerging from a common spacelike 2-disk can
serve as a Cauchy surface for a region of spacetime. Already in the 1960s free
(unconstrained) initial data for general relativity were found for such
hypersurfaces. Here an expression is obtained for the symplectic 2-form of vacuum
general relativity in terms of such free data. This can be done, even though variations
of the geometry do not in general preserve the nullness of the initial
hypersurface, because of the diffeomorphism gauge invariance of general relativity.
The present expression for the symplectic 2-form has been used previously \cite{PRL}
to calculate the Poisson brackets of the free data.
\end{abstract}




\section{Introduction} 

\begin{figure}
\begin{center}

\includegraphics{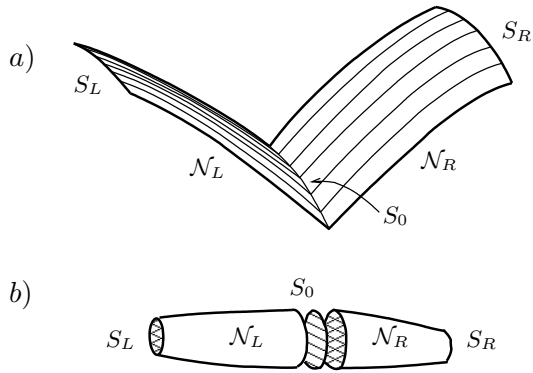}
\caption{a) A double null sheet in 2+1 dimensional spacetime. b) In 3+1 
dimensional spacetime $\cN$ is a 3-manifold consisting of two solid cylinders
joined on a disk (shown here without regard to their embedding in spacetime).}
\label{Nfigure}
\end{center}
\end{figure}

Free (unconstrained) initial data for General Relativity (GR) on certain 
piecewise null hypersurfaces have been known since the 1960s 
\cite{Sachs,Dautcourt,Penrose}. In the present work the symplectic 
2-form corresponding to the Einstein-Hilbert action for vacuum GR is expressed
in terms of such free data on a so called {\em double null sheet}, a compact 
hypersurface $\cN$, consisting of two null branches, $\cN_L$ and $\cN_R$, that 
meet on a spacelike 2-disk $S_0$ as shown in Fig.\ \ref{Nfigure}.\footnote{
Some of this work has been reported in the e-print \cite{MR} and in the
letter \cite{PRL}, where the symplectic 2-form is used to obtain a
Poisson bracket on the free null initial data.}
$\cN_L$ and $\cN_R$ are swept out by the two 
congruences of future null normal geodesics (called {\em generators}) emerging 
from $S_0$, and are truncated on disks $S_L$ and  $S_R$ respectively before 
the generators form caustics.\footnote{
Caustic points are points where the generators ``focus"; Roughly speaking, 
where neighboring generators meet. More precisely, they are points where the
differences in the coordinates of points of equal parameter value on 
neighboring generators vanishes to first order in the differences of 
coordinates of the base points of the two generators at $S_0$. This does 
not quite imply that the generators actually meet. }
With this symplectic 2-form the space of valuations of the free data becomes 
a phase space, which, among other things, may serve as a starting point for 
quantization.

In most initial value formulations of GR the initial data is subject to 
constraints, which complicates canonical formulations based on those data.
In fact, at present the handling of the constraints absorbs most of the effort 
invested in canonical approches to quantum gravity. A canonical formulation 
based on {\em } free initial data is thus of considerable interest. 

To be sure, null data is not the only way to obtain a constraint free canonical 
theory. York \cite{York72} has identified spacelike free initial 
data, and has set up a canonical theory on spacelike hypersurfaces of 
uniform mean extrinsic curvature in which the most difficult constraint, the 
scalar constraint, has been eliminated (see \cite{Bruhat_York}).


A canonical framework based on null hypersurfaces is, however, especially 
suited for addressing certain issues. In particular the canonical framework 
obtained here and in \cite{PRL} seems ideal for attempting a semi-classical
proof of Bousso's formulation of the holographic entropy bound
\cite{Beckenstein,tHooft,Susskind,Bousso} in the vacuum gravity case, since a 
branch $\cN_A$ ($A = L\ \mbox{or}\ R$) of $\cN$ is a ``light sheet'' in the 
terminology of Bousso \cite{Bousso} (provided the generators are not  
expanding at $S_0$). It also seems a good classical starting point for a 
search for a quantization of GR respecting this entropy bound. That is, a 
quantization in which the area of $S_0$ has a discrete spectrum 
and each eigensubspace is of finite dimension, bounded by the exponential of 
the maximal entropy, according to Bousso's bound, of $\cN_L$ and $\cN_R$ 
together, i.e. by $exp(\mbox{Area($S_0$)}/2 \mbox{Planck area})$.\footnote{
In order that Bousso's bound apply to both branches of $\cN$ the generators
on both sides of $S_0$ must be non-expanding. (This does not imply that
$S_0$ lies in a black hole, for $S_0$ is a disk, not a boundaryless closed
surface.) Such $S_0$ are easily constructed even in flat spacetime: For example,
take $S_0$ to be a portion of the intersection of two past light cones.}


In \cite{PRL} and the preprint \cite{MR} the symplectic 2-form,
$\omega_\cN$, was used to calculate the Poisson brackets between initial data 
on $\cN$. The main aim of the present work is to provide a detailed derivation 
of the expression for $\omega_\cN$ that was used.
The symplectic 2-form at a solution metric $g$ takes as arguments two variations
$\dg_1$ and $\dg_2$ belonging to the space $L_g$ of smooth solutions to the field
equations linearized about $g$.
The expression for $\omega_\cN[\dg_1,\dg_2]$ in terms of free null initial data
obtained here is valid for all ``admissible'' $\dg_1$ and $\dg_2$. Admissible
variations preserve the null character of the branches of $\cN$ and some other
structures associated with $\cN$. Because of diffeomorphism gauge invariance the 
expression also holds in an slightly indirect way for a much larger class of variations.
If $\dg_1, \dg_2 \in L_g$ and $\dg_2 g_{ab}$ vanishes in a spacetime neighborhood
of $\di\cN$, then there exist corresponding admissible variations $\dg'_1$ and $\dg'_2$
such that $\omega_\cN[\dg_1,\dg_2] = \omega_\cN[\dg'_1,\dg'_2]$. The symplectic product
$\omega_\cN[\dg_1,\dg_2]$ may therefore be expressed in terms of the variations of the
free null initial data under $\dg'_1$ and $\dg'_2$. This suffices to obtain a Poisson
bracket between the initial data.

To understand this let us briefly review how the Poisson bracket is obtained in 
\cite{PRL}: On a finite dimensional phase space with non-degenerate symplectic 
2-form\footnote{
In the present work the requierment of non-degeneracy is {\em not} part of the
definition of a symplectic 2-form. Both symplectic and presymplectic 2-forms are
referred to as ``symplectic 2-forms''. This is convenient because whether or not
the symplectic 2-form is degenerate depends on the set of variations one admits,
and on a given space of variations whether it is degenerate is generally not 
obvious {\em a priori}.}
the Poisson bracket is determined by the inverse of this 2-form. In the
case of initial data for general relativity on $\cN$ subtleties arise, both 
because $\cN$ has boundaries, and because the data has infinitely many degrees
of freedom. In an infinite dimensional phase space a non-degenerate symplectic
2-form can fail to have an inverse because it does not map onto the whole
covector space. This is the case here. The inverse of the symplectic 2-form
does not define Poisson brackets between all modes of the initial data.

This lead the author to look for a new starting point.
The Peierls bracket \cite{Peierls} is an alternative expression for the Poisson 
bracket which does not depend directly on the symplectic 2-form. The Peierls bracket 
between two functionals of spacetime fields is given by a very simple expression 
in terms of the first order perturbations to the solutions of the field equations 
occasioned by adding these functionals to the action. It's simplicity, and its 
direct relation to the quantum commutator give it a good claim to being a more 
fundamental definition of the Poisson bracket than the one in terms of the 
symplectic 2-form. Furthermore it agrees with the latter definition when both are 
defined \cite{Peierls, DeWitt, MR}.
  
Unfortunately the Peierls bracket between data on $\cN$ is ambiguous, because 
the perturbation generated by a functional of data on a characteristic 
hypersurface is discontinuous precisely at the hypersurface itself. The Peierls 
bracket {\em is} well defined on so called "observables", diffeomorphism invariant
functionals $F[g]$ of the metric, with smooth functional derivatives $\dg F/\dg g_{ab}$
of compact support contained in the interior of the causal domain of dependence of
$\cN$.\footnote{
The causal domain of dependence $D[S]$ of a set $S$ in a Lorentzian signature 
spacetime is the set of all points $p$ such that every inextendible causal curve 
through $p$ intersects $S$. If $S$ is a closed achronal hypersurface one expects
in physical theories that initial data on $S$ fixes the solution in $D[S]$. 
See \cite{Wald}.}\footnote{
In \cite{MR} a wide class of examples of observables in this
sense is constructed, which determines the spacetime geometry of the domain of 
dependence, at least for generic geometries.}

The approach of \cite{PRL, MR} is to look for a Poisson bracket
$\{\cdot,\cdot\}_\bullet$ on initial data that reproduces the Peierls brackets 
between observables. In \cite{MR} it is shown that to ensure this match between
the $\bullet$ bracket and the Peierls bracket (in a spacetime with metric $g$ satisfying
the field equations) it is sufficient to require that
\be                     \label{auxbracketdef}
        \dg A = \omega_\cN[\{A,\cdot\}_\bullet,\dg],
\ee
for any observable $A$ and any $\dg$ in the space $L_g^0$ of smooth variations which satisfy
the field equations linearized about $g$ and vanish in a spacetime neighborhood of $\di\cN$.

When both sides of (\ref{auxbracketdef}) are expressed in terms of the initial 
data on $\cN$ it becomes a condition on the Poisson brackets of these data.
(In fact this condition is nothing but a suitably weakened form of the requirement
that the Poisson bracket be inverse to the symplectic 2-form.) To express
(\ref{auxbracketdef}) in terms of initial data $\omega_\cN[\dg_1,\dg_2]$ must
be expressed in terms of the initial data, but only in the case that $\dg_2$
vanishes in some neighborhood of $\di\cN$.


Sachs \cite{Sachs} and Dautcourt \cite{Dautcourt} showed formally that any
valuation of their null initial data on $\cN$ determines a matching solution
which is unique up to diffeomorphisms. This is the basis of their claim that
their data, which is equivalent to the data we will use, is free and complete.
And, of course it is the basis of the program of canonical general relativity
in terms of these null initial data.
Because their analyses do not address convergence issues they do not give a
clear indication of the domain on which the solution exists or is unique.
It seems reasonable to expect that the data in fact determine a maximal
Cauchy development of $\cN$, but what has been demonstrated rigorously so
far is that a solution matching the data exists and is unique in some
neighborhood of $S_0$ in the future of $\cN$ \cite{Rendall}. It has not
been established that there is always a development of {\em all} of $\cN$.

It is therefore worth noting that the existence and uniqueness of Cauchy developments of the data
is not strictly necessary for the results of the present work. The space of data, the 
symplectic 2-form, and the Poisson bracket on the data found in \cite{PRL}, are all defined 
independently of Cauchy developments. Indeed it is possible, and perhaps fruitful, to 
define a phase space of initial data on just a single branch of $\cN$, even though the 
data on a single branch cannot by itself define a Cauchy development.  



Given that free null initial data for GR has been available for such a long 
time the question arises as to why a canonical framework based on such data was 
not developed sooner. In fact canonical GR using {\em constrained data} on 
double null sheets has been developed by several researchers \cite{Torre, 
Goldberg1, Goldberg2, Vickers}. Also, partial results have been obtained on the 
Poisson brackets of free data \cite{GR, Goldberg2}. In \cite{GR} Gambini and 
Restuccia give perturbation series in Newton's constant for the brackets of free 
data living on the bulk of $\cN$, but no brackets for other (necessary) data 
that live on the intersection surface $S_0$. Their results are consistent with 
the present work and were indeed crucial for its genesis.\footnote{
The brackets 
between the bulk data given by the author in \cite{PRL} were first obtained
by summing the series of Gambini and Restuccia in closed form and simplifying 
the result by a change of variables, before being derived more systematically 
from the symplectic 2-form obtained here (and in \cite{MR}).}
In \cite{Goldberg2} Goldberg and Soteriou present distinct free data 
on the bulk of $\cN$, which are claimed to form a canonically conjugate pair 
on the basis of a machine calculation of their Dirac brackets. It would be 
interesting to see if they are conjugate according to the symplectic 
structure obtained here.


There is however a conceptual issue which seems to have discouraged many
researchers from trying to develop null canonical theory. Namely the problem
of generator crossings and caustics. This problem is actually much less serious 
than it seems.


Let us briefly examine the problem, and its solution. Although it is not
relevant to the main task of the present paper, which is to evaluate the 
symplectic 2-form in terms of free initial data, it is relevant to the 
viability of the over-all program of developing a canonical formulation of 
GR based on these free null initial data.


\begin{figure}
\centering
\scalebox{0.77}{\input{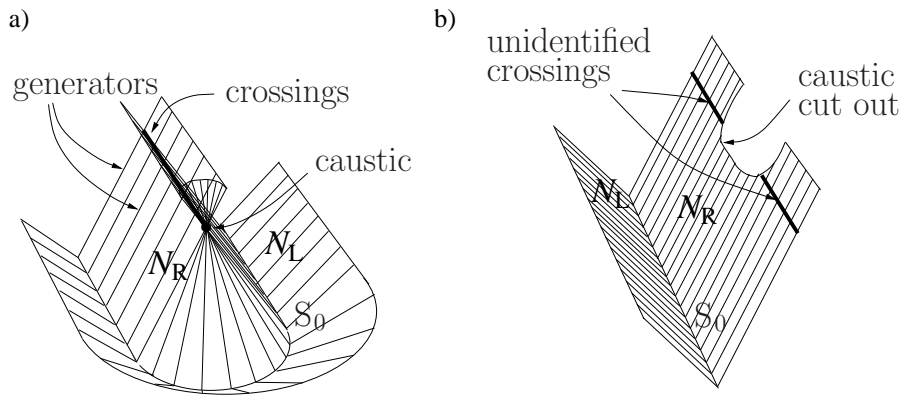}}
\caption{Panel a) shows a simple example of a caustic and intersections of 
generators in $2 + 1$ Minkowski space: $S_0$ is a spacelike curve having the 
shape of a half racetrack - a semicircle extended at each end by a tangent 
straight line. The congruence of null geodesics normal to $S_0$ and directed 
inward and to the future sweep out $\cN_R$, which takes the form of a ridge roof,
terminated by a half cone over the semicircle. The generators from the 
semicircle form a caustic at the vertex of the cone. There neighbouring 
generators intersect. On the other hand generators from the two straight 
segments of $S_0$ cross on a line (the ridge of the roof) starting at the 
caustic, but the generators that cross there are not neighbours at $S_0$.
Clearly the generator segments beyond the crossing points enter the interior
of the domain of dependence of $\cN$.
In Panel b) the double null sheet defined by $S_0$ in the covering space
is shown, with the points that are
identified in the original spacetime indicated.}
\label{crossing}
\end{figure}

The problem is the following: Suppose a double null sheet $\cN$ is 
constructed in a given solution spacetime $M$. It can easily happen 
that the generators that sweep out $\cN$ pass through a caustic and/or cross 
if extended far enough. See Fig.\, \ref{crossing}. Once this occurs the 
generators enter the chronological future of $\cN$ (see \cite{Wald} Theorem 
9.3.8). In fact the segments of the generators beyond caustic or crossing 
points enter the interior of the domain of dependence of $\cN$. The portion of 
$\cN$ composed of these segments lies in the domain of dependence of the 
remainder of $\cN$. See Appendix B of \cite{MR} 
The initial data on
part of $\cN$ will thus be determined by the solution defined by the data on 
the rest of $\cN$, which constitutes a highly complex constraint on data 
which was supposed to be free.\footnote{
This argument supposes that the solution matching the data is unique on the
whole domain of dependence, which has not been established. However the
solutions to the linearized field equations are certainly unique on this
domain, and this already precludes independent continuous variations of
the data on the part of $\cN$ lying in the interior of the domain of 
dependence.}

Thus one would apparently wish to exclude initial data corresponding to 
hypersurfaces containing caustics and crossings from the phase space. A 
condition excluding caustics is easily found, but it seems to be much more 
difficult to exclude non-caustic crossing points. Presumably one would have 
to impose some sort of non-local inequality, which would also rob the phase 
space of free initial data of its simplicity.


In fact this is unneccesary. Once caustics have been excluded from $\cN$ any 
further crossing points can be ``unidentified" because there exists an
isometric covering spacetime in which the generators do not cross, formed by
pulling the metric back to the normal bundle of $S_0$ via the exponential map.
(See Appendix B of \cite{MR}.) In this new spacetime no constraint
forbids the independent variation of the free initial data on all parts of 
$\cN$. Of course once the data is changed there is no guarantee that the
spacetime regions that were unidentified in going to the covering spacetime are 
still isometric, so it may no longer be possible to identify them. The 
complicated "constraints" arising from generator crossings in the original 
spacetime are precisely the conditions that must be met in order that the 
isometry of these regions be maintained. They are {\em not} constraints that 
must be satisfied in order that a solution matching the data exists.

We are thus led to the following simple and plausible picture: Any valuation
of the free data without caustics on $\cN$ posseses a Cauchy development
satisfying Einstein's equations. The Cauchy developments of a subset of 
valuations of the initial data, which satisfy certain complicated conditions, 
have isometries which allow the identification of regions so that the 
generators of $\cN$ cross in the resulting spacetime.\footnote{
It is worth noting that the same issue arises in the spacelike Cauchy 
problem, and is resolved in the same way. Spacelike hypersurfaces that enter 
the interior of their own domains of dependence are easily constructed in 
any solution spacetime $M$. But the unique maximal Cauchy development of the 
initial data induced from $M$ on such a hypersurface is a covering manifold 
of the original domain of dependence, in which the hypersurface is 
achronal.}    
Note that we have not proved that this picture is correct. That requiers a 
proof of the existence of solutions matching the free data throughout $\cN$, 
which is not yet available. What has been shown is that the possibility of 
generator crossings does not represent an obstruction to this picture, nor 
even an argument against it.  


This resolution of the problem of generator crossings suffices for the 
development of a simple and meaningful canonical theory based on null initial 
data. However, it does not mean that generator crossings are always to be 
regarded as unphysical. In many applications one surely would have to deal with 
them. But even in such cases a canonical framework based on Cauchy developments 
in which all generator crossings have been unidentified might provide a useful 
perspective.  


A different conceptual issue, which {\em is} directly relevant to the 
calculation of the symplectic 2-form, is the following: The symplectic 2-form 
is a bilinear on perturbations of the metric satisfying the linearized field 
equations. Generically such perturbations do not preserve the null character of 
the branches of $\cN$. How then are these perturbations to be represented by the variations of
{\em null} initial data? The key is the diffeomorphism gauge invariance of GR.
Roughly speaking, to each perturbation there corresponds a gauge equivalent one
which does preserve the nullness of the branches of $\cN$, and so can be expressed in
terms of the variation of null initial data. This is only aproximately correct. 
As we will see, the precise resolution of the problem is rather delicate because
not all diffeomorphisms are gauge in the sense of being degeneracy vectors of 
the symplectic 2-form.  
 

The remainder of the paper is organized as follows: In the next section 
the free initial data that will be used is defined using a convenient chart on 
each of the branches of $\cN$. This data is shown to be equivalent to Sachs's
data, and thus free and complete to the extent that Sachs's is. In section 
\ref{symp1} the symplectic 2-form corresponding to the Einstein-Hilbert action
is evaluated on an arbitrary hypersurface in terms of the 4-metric and its variations.
A large class of infinitesimal diffeomorphisms is shown to be gauge in subsection \ref{diffeos}.
Section \ref{symp_null} is dedicated to expressing the symplectic 2-form in terms of
the null initial data. In subsection \ref{expressing_with_null} it is shown how the
diffeomorphism gauge invariance can be exploited to express the symplectic 2-form in
terms of null initial data on the variations that needed for the calculation of the
Poisson bracket in \cite{PRL}. Subsection \ref{role_of_diff_data} is a discussion of the
role of the diffeomorphism data. In subsection \ref{charts} some important charts are
defined. In subsection \ref{symp_pot} the symplectic potential is expressed in terms
of our free the null initial data. Finally, in subsection \ref{symp2} the symplectic
2-form is obtained in terms of these data.
An appendix treats variations in fixed and moving charts.

\section{The free data}

\subsection{coordinates on $\cN$}\label{coordinates}

A special chart $(v^A, \theta^1, \theta^2)$ will be used on each branch $\cN_A$ 
($A = L\ \mbox{or}\ R$) of $\cN$. $v^A$ is a parameter along the generators 
and $\theta^p$ ($p = 1,2$) is constant along these. Since $\di_{v^A}$ is tangent 
to the generators it is null and normal to $\cN_A$.\footnote{
{\em Proof}: Suppose $t$ is tangent to $\cN_A$ at $p \in \cN_A$, 
then $t$ may be Lie dragged to $S_0$ along $n_A \equiv \di_{v^A}$, staying 
always tangent to $\cN_A$, and
\be \label{diff_eq}
\di_{v^A} (n_A \cdot t) = [\nabla_{n_A} t]\cdot n_A + [\nabla_{n_A} n_A]\cdot t.
\ee
The first term vanishes since $[\nabla_{n_A} t]\cdot n_A = [\nabla_t n_A]\cdot n_A
= 1/2\nabla_t n_A^2 = 0$. Because the generators are geodesics the second term 
reduces to $\ag n_A \cdot t$ with $\ag$ a scalar measuring the non-affineness 
(i.e. acceleration) of the parameter $v^A$.  $n_A \cdot t$ vanishes at $S_0$, 
since there $t$ is a sum of tangents to $S_0$ and $n_A$, which are both normal 
to $n_A$, the null normal to $S_0$. (\ref{diff_eq}) then shows that it vanishes 
also at $p$.\QED}
The line element on $\cN_A$ thus takes the form
\be   \label{line_element}
ds^2 = h_{pq} d\theta^p d\theta^q,
\ee
with no $dv$ terms. $v^A$ is taken proportional to the square root of 
$\rho \equiv \sqrt{\det h}$, the area density in $\theta$ coordinates on 2D 
cross sections of $\cN_A$, and normalized to $1$ at $S_0$. Thus 
$\rho = \rho_0(\theta^1, \theta^2) v^2$, with $\rho_0$ the area density on 
$S_0$. $v$ will be called the {\em area parameter}.\footnote{
The index $A$ specifying the branch $\cN_A$ of $\cN$ will often be dropped 
when there is no risk of confusion.} 

The area parameter is related to affine parameters on the generators by the
vacuum Einstein equation contracted with the tangents of the generators,
$R_{vv} \equiv R[\di_v, \di_v] = 0$. Suppose $\eta$ is an affine parameter
along the generators of $\cN_A$. Then, because of this field equation and because
the generators are surface forming, the Raychaudhuri equation (\cite{Wald},
eq. (9.2.32)) reduces to the focusing equation
\be \label{focusing0}
\frac{d\theta}{d\eta} = -\frac{1}{2}\theta^2 - \sg_{pq}\sg^{pq},
\ee
where $\theta$ is the expansion, and $\sg$ is the shear. Now (see \cite{Wald} eq. (9.2.28))
\be
h_{pq}\,\theta  + 2\sg_{pq} = {\pounds}_k h_{pq} = \di_\eta h_{pq} = \rho\, \di_\eta e_{pq} + h_{pq}\,\di_\eta \ln \rho ,
\ee
where $k = \di_\eta$ is the $\eta$ tangent to the generators, and the partial derivatives
are evaluated in the chart $(\eta, \theta^1, \theta^2)$. 
It follows that the expansion is $\theta = \di_\eta \ln \rho = 2\di_\eta \ln v$, that the shear is
$\sg_{pq} = \rho/2\, \di_\eta e_{pq}$, and that
\be
\sg^{pq} = h^{ps}h^{qt}\sg_{st} = \frac{1}{\rho^2} e^{ps}e^{qt}\sg_{st} = -\frac{1}{2\rho}\di_\eta e^{pq}.
\ee
Substituting these expressions into (\ref{focusing0}) one finds
\be
\di_\eta^2 \ln v + (\di_\eta \ln v)^2 = \frac{1}{8} \di_\eta e_{pq} \di_\eta e^{pq}.
\ee 
Finally, changing the variable of differentiation to $v$ we obtain
\be \label{focusing}
\frac{d}{dv}\ln \Big|\frac{d\eta}{dv}\Big| = - \frac{v}{8}\di_v e_{pq}\di_v e^{pq}.
\ee 
This is the key equation that relates our free initial data to Sachs' \cite{Sachs} free initial data.

Using $v$ as a coordinate makes avoiding caustics easy. At caustic points 
$v^2 \equiv \rho/\rho_0$ vanishes, so the caustic {\em free} $\cN$ are 
represented by initial data on coordinate domains in which $v >0$.

On the other hand, $v$ is not always a good parameter on the generators. 
For instance it fails in the important special case in which $\cN_A$ is 
a null hyperplane in Minkowski space, because the generators neither 
converge nor diverge, resulting in a $v$ that is constant on each 
generator. Nevertheless, for generic $\cN$ in generic spacetimes $v$ is 
good enough. Indeed in the case of greatest interest from the point of 
view of the holographic entropy bound, in which the generators are 
converging everywhere on $S_0$ ($v$ decreasing away from $S_0$), 
the focusing equation (\ref{focusing}) ensures that $v$ continues to 
decrease until a caustic is reached. Since the generator segments in $\cN$
are truncated before reaching a caustic this implies that $v$ is a good 
parameter on $\cN$. 

The area parameter $v$ is also a good parameter if the generators are diverging at $S_0$, 
provided they are truncated before they begin to reconverge. If the 
generators converge on some parts of $S_0$ and diverge on others our
methods may still be used. Suppose $p$ is a point on $S_0$ at which the 
expansion of both the $R$ and the $L$ future null normals is non-zero (and 
suppose both the spacetime geometry and $S_0$ are smooth\footnote{
A smooth function on a domain with boundary is {\em defined} to be one that posses a
smooth extension to an open domain. See \cite{AMR} chapter 7.
Consequently a smooth manifold with boundary necessarily has an extension to a 
smooth manifold without boundary, and an embedding of a manifold with boundary 
is smooth iff there exists a smooth extension of the embedding to a manifold 
without boundary}
), then this will also be true throughout a small disk 
$S'_0 \subset S_0$ about $p$. The chart $(v^A, \theta^q)$ is thus good on each
branch of a double null sheet $\cN' \subset \cN$ swept out by the generators 
emerging from $S'_0$. The symplectic 2-form may thus be computed on $\cN'$, and 
from it the Poisson brackets between the data on $\cN'$. Causality requires 
that these in fact be all the non-zero Poisson bracket of the data on the 
generators through $p$. The only points of $\cN$ that are causally connected to 
a point on these generators are the points of these generators themselves, all 
others are "spacelike separated" from them (see appendix B of \cite{MR}), so data 
on distinct generators should have vanishing Poisson brackets. Indeed this is what 
is found when the brackets are computed \cite{PRL}.\footnote{
These Poisson brackets are calculated {\em assuming} that $v^A$ is a good 
parameter on the generators throughout $\cN_A$. That is, they are the brackets
between data on $\cN'\subset \cN$ and not necessarily all of $\cN$. This is 
therefore not a proof that data on all distinct generators Poisson commute, 
but it does mean that the brackets that could be calculated are consistent 
with this expectation coming from causality.}

In the following we shall assume, without great loss of generality according
to the preceeding arguments, that $v$ is a good parameter throughout each
branch of $\cN$.

Ultimately, in order to define a phase space of the gravitational field
in terms of initial data we have to express all limitations on admissible
solutions and coordinates as restrictions on the initial data (expressed as 
functions of the coordinates). Points at which the parameter $v$ is 
stationary, and thus not a good parameter, turn out to be detectable in
the field $e_{pq}$ on $\cN$, which will be one of our data. 
Integration of (\ref{focusing}) yields
\be \label{int_focusing1}
\frac{dv}{d\eta}(v) 
= \frac{dv}{d\eta}(v_0)\exp{\int_{v_0}^v \frac{v}{8}\di_v e^{pq} 
\di_v e_{pq} dv}. 
\ee
Since $d/d\eta$ is the parallel transport of a non-zero vector at $S_0$,
it is non-zero everywhere on the generator, so (\ref{int_focusing1}) implies
that 
\be
dv|_v = dv|_{v_0}\exp{\int_{v_0}^v \frac{v}{8}\di_v e^{pq} \di_v e_{pq} dv}, 
\ee
along the generators. Therefore if $v$ is a good parameter ($dv \neq 0$ on the 
generator) at {\em some} value $v_0$, and $e_{pq}$ is a continuously 
differentiable function of $v$ then $v$ is a good parameter at all 
finite values of $v$. A breakdown of $v$ as a parameter requires a (sufficiently 
strong) singularity in $\di_v e_{pq}$. We shall admit only initial data that 
is smooth in the coordinates, so $v$ is guaranteed to be good.

On a branch $\cN_A$ the coordinate $v^A$ thus ranges from $1$ on $S_0$ to
its value, $\bar{v}^A$, on $S_A$, $\bar{v}^A$ being a smooth function of the 
$\theta$ which is $>0$ and $\neq 1$. 

\subsection{The data}\label{data}

Two types of data will be used: geometrical data that reflect the spacetime
geometry, that is the diffeomorphism equivalence class of the metric, and
diffeomorphism data which reflect the choice of metric within the diffeomorphism
equivalence class.

The inclusion of the diffeomorphism data may seem odd in a diffeomorphism invariant theory. However
the geometrical data are not enough to express the symplectic 2-form on $\cN$
for all the variations we will consider. Because $\cN$ has a boundary, not all
infinitesimal diffeomorphisms are degeneracy vectors of the symplectic 2-form,
$\omega_\cN$, on $\cN$. That is, some degrees of freedom measuring diffeomorphisms
of the spacetime metric are non-gauge in the sense that their variations contribute
to the symplectic 2-form. In order to be able to express the symplectic 2-form
in terms of the variations of initial data on $\cN$ it is therefore necessary
in general to include in the data variables parametrizing these degrees of freedom.
This does not necessarily mean that the diffeomorphism data are ``physical''.
Indeed they seem to play no essential role in the phase space formulation
of vacuum general relativity within the domain of dependence of $\cN$. They seem
rather to be auxiliary quantities used in the intermediate stages of the construction
of this formulation. They may however be important for the definition of quasi-local
linear and angular momenta associated with $\cN$.

The diffeomorphism data will be discussed at the end of this section.
The geometrical data we will use consist of $e_{pq}$,
specified on the branches of $\cN$ as a function of the $v$ and $\theta$
coordinates, and further data given only on $S_0$ as functions of the
$\theta^p$, namely $\rho_0$, $\lam = - \ln|n_L\cdot n_R|$, and the
{\em twist}
\be	\label{twist}
\tau_p = \frac{n_L\cdot\nabla_p n_R - n_R\cdot\nabla_p n_L}{n_L\cdot n_R}.
\ee
Here $n_A = \di_{v^A}$ is the tangent to the generators of $\cN_A$, and
inner products ($\cdot$) are taken with respect to the spacetime metric.
These data will be called {\em $v$ data}. 
They are {\em regular} if the data on $S_0$ are smooth functions of the $\theta$ chart,
and $e_{pq}$ is smooth in the $v\theta$ chart on each branch of $\cN$, as well as
continuous across $S_0$.

Smooth solutions induce regular $v$ data on any smooth double null
sheet $\cN$, provided that on each branch the generators are either everywhere
converging or everywhere diverging and free of caustics, and the $\theta^p$
form a smooth chart on $S_0$. (When the generators are everywhere converging
or diverging, $v$ is a smooth function without stationary points on the generators.
Smooth functions on the generators are then smooth functions of $v$.)

Sachs \cite{Sachs} argues that a similar set of data is free, and complete in
the sense that it 
determines the solution geometry. Sachs' data consists of $e_{pq}$ on $\cN$,
but given as a function of an affine parameters $\eta$ on the generators
instead of $v$, and the following data on $S_0$: $\rho_0$, 
$\di_{\eta^L} \rho$, $\di_{\eta^R} \rho$, and $\tau_\eta$ (which is the
twist (\ref{twist}), but calculated from the tangents $\di_{\eta^A}$ instead
of the $n_A = \di_{v^A}$).\footnote{
Sachs actually takes as his final datum a pair of quantities he writes as 
$C_{A,1}\ \ \ A = 1,2$. These are in fact the components of $-\tau_\eta$, as 
can be seen most easily from his equation 19. When a forgotten factor 
of $1/2$ is restored and it is rewritten in our notation this equation reads
\be
\frac{1}{2} C_{p,1} = \di_{\eta^L} \cdot \nabla_p \di_{\eta^R}.
\ee
The normalization condition $\di_{\eta^L} \cdot \di_{\eta^R} = -1$, which 
Sachs imposes on the affine parameters, implies that the right side equals 
$-1/2\tau_{\eta\,p}$.}

Regular $v$ data is equivalent to Sachs data. We will demonstrate that all regular $v$
data determine unique corresponding Sachs data such that any solution matching the $v$
data also matches the Sachs data, and conversely, any solution matching the Sachs data
matches the original $v$.
It follows that if the Sachs data is free and complete, then regular $v$ data is also:
Suppose a solution matches a set of $v$ data, then it also matches a uniquely determined
set of Sachs data. If the Sachs data determines the solution uniquely (up to diffeomorphisms)
then so does the $v$ data. That is, the $v$ data is complete. To establish that it is free
it must be shown that any regular $v$ data matches a solution. But if Sachs data are free
then the Sachs data corresponding to the $v$ data necessarily match a solution, and
this solution also matches the $v$ data.

In fact it has been proved by Rendall that any smooth Sachs data\footnote{
In his proof of existence and uniqueness Rendall takes as a datum $\di_L g_{Rp}$ 
(where $g$ is the 4-metric and the components are referred to the basis 
$d\eta^L, d\eta^R, d\theta^p$) in place of $\tau_{\eta\,p} = 2\di_{[L} g_{R]p}$.
But in Rendall's spacetime coordinates $\di_{(L} g_{R)p}$ is determined by
the remaining (Sachs) data, so his proof applies just as well if $\tau_\eta$
is used as the datum.}
matches a unique solution {\em in some neighborhood of $S_0$} 
\cite{Rendall}, and it is a reasonable conjecture that it matches a unique 
solution on all of $\cN$ provided $\cN$ is free of caustics. (See 
discussion in the introduction.) The Sachs data corresponding to regular 
$v$ data are indeed free of caustics on $\cN$. Thus, if the conjecture is 
valid, regular $v$ data are free and complete on $\cN$.

We now turn to the proof of the equivalence of regular $v$ data and Sachs data.
The proof consists in demonstrating that in solution spacetimes regular $v$
data on $\cN$ determines the Sachs data on $\cN$. Moreover, without assuming
{\em a priori} that a solution matching the $v$ data exists, Sachs data may be
evaluated for {\em any} regular $v$ data using the transformation that holds on
solutions. Finally, it is noted that any solution matching Sachs data obtained
in this way from regular $v$ data also matches the original $v$ data.

As already mentioned, the Sachs data differ from the $v$ data essentially by
a coordinate transformation. The Sachs data are functions of an affine parameter
along the generators, while the $v$ data are functions of the area parameter $v$.
As the first step in the equivalence
proof let us demonstrate that in a solution an affine parameter $\eta$ along the
generators can be calculated from the $v$ data and the area parameter $v$. $\eta(v)$
then determines the map from the coordinates $v$, $\theta^1$, $\theta^2$, to
which the $v$ data are referred, to Sachs' coordinates $\eta$, $\theta^1$, $\theta^2$.

The field equation $R_{vv} = 0$ on $\cN$ implies 
that any affine parameter $\eta$ along the generators satisfies the focusing
equation (\ref{focusing}). But from the integrated form (\ref{int_focusing1})
of the focusing equation it is clear that $e_{pq}$, which is a smooth function of
$v$ on the compact interval $[1, \bar{v}]$, determines $\eta(v)$ up to an 
affine transformation, that is, up to a constant rescaling and a constant shift. 
The solutions to (\ref{int_focusing1}) are thus precisely the affine 
parameters.\footnote{
When the field equation $R_{vv} = 0$ does not hold $\eta(v)$ is not an
affine parameter, but it is still determined up to affine transformations, 
and it is a good parameter, for it follows directly from 
(\ref{int_focusing1}) that $\eta(v)$ is smooth and monotonic with non-zero
derivative. If the parameter $\eta$ that Sachs data is referred to is 
interpreted to be this parameter then all spacetime geometries, solutions or 
not, that match regular $v$ data, also match the corresponding Sachs
data. The only role of the field equations in the equivalence of regular $v$
data and Sachs data is that they ensure that $\eta(v)$ is an affine parameter
in accordance with the standard spacetime interpretation of Sachs data.}

The shift and rescaling freedom in $\eta(v)$ can be parameterized by the 
values of $\eta$ and $\di_v\eta$ at $S_0$. For the Sachs coordinates
$\eta^L$ and $\eta^R$ this amounts to four functions $A_A = \eta^A|_{S_0}$ 
and $B_A = \di_{v^A} \eta^A|_{S_0}$ on $S_0$, which are restricted by Sachs´ 
condition $\di_{\eta^L} \cdot \di_{\eta^R} = -1$ at $S_0$. Rewriting this 
condition in terms of the vectors $n_A \equiv \di_{v^A} = B_A \di_{\eta^A}$
one obtains
\be
- B_L B_R  = n_L \cdot n_R 
= - \sg_L \sg_R |n_L \cdot n_R|= - \sg_L \sg_R e^{-\lam},
\ee
where $\sg_A = 1$ if $v^A$ increase toward the future (i.e. $\bar{v}_A > 1$), 
and $\sg_A = -1$ if it decreases toward the future. (The signature of the 
4-metric is taken to be ${}-+++$, which implies that the inner product of future 
directed tangents to the $L$ and $R$ generators is negative). 

Since $e^{pq}$ is smooth in $v$ and $v$ is non-stationary on the generators
(\ref{int_focusing1}) implies that $\eta(v)$ is smooth, with non-zero derivative,
and thus has a smooth inverse $v(\eta)$. As claimed, the $v$ data (and the
parameters $A_L$, $A_R$, and $B_R$ or $B_L$) determine a smooth and smoothly
invertible transformation from the chart $(v^A, \theta^p)$ to Sachs' chart
$(\eta^A, \theta^p)$.
The shift and rescaling degrees of freedom can be eliminated
by fixing the parameters $A_L$, $A_R$, and $B_R$ once and for all.
We will set $A_L = A_R = 0$, and $B_R = 1$. 

The coordinate transformation allows us to obtain $e_{pq}$ as a function
Sachs coordinates. Note that $e_{pq}$ transforms as a scalar under this 
particular change of chart. This is because the line element on $\cN_A$,
$ds^2 = h_{pq} d\theta^p d\theta^q$, is degenerate, with no contribution 
from displacements along the generators. Since the $\theta$ coordinates
are the same in the two charts the components $h_{pq}$ at a given point
on $\cN_A$ are the same. That is, $h_{pq}$ transforms as a scalar under
the change of charts. It follows that $e_{pq} = h_{pq}/\sqrt{h}$ does also.
The result of the transformation, $e_{pq}(\eta, \theta^1, \theta^2)$, is
smooth, and continous across $S_0$.

It remains to calculate the Sachs data on $S_0$. Namely $\rho_0$, 
$\di_{\eta^L} \rho$, $\di_{\eta^R} \rho$, and $\tau_\eta$. 
The $v$ data of course already includes $\rho_0$, and the derivatives of
$\rho = \rho_0 v^2$ are easily obtained from the $v$ data: 
\bearr
\di_{\eta^R} \rho|_{S_0} & = & \rho_0 2 v^R \di_{\eta^R} v^R|_{S_0} 
= 2\rho_0 B_R^{-1} = 2\rho_0,   \label{drho_R}\\
\di_{\eta^L} \rho|_{S_0} & = & \rho_0 2 v^L \di_{\eta^L} v^L|_{S_0}
= 2\rho_0 \sg_L\sg_R e^\lam B_R = 2\rho_0 \sg_L\sg_R e^\lam.  \label{drho_L}
\eearr
Finally, $\tau_{\eta\, p}$ is given by the same expression as the $v$ datum
$\tau_p = [n_L\cdot\nabla_p n_R - n_R\cdot\nabla_p n_L]/n_L\cdot n_R$,
but with the vectors $n_A$ substituted by $\di_{\eta^A} = B_A^{-1} n_A$.
Thus
\bearr	
\tau_\eta & = & \tau + d \ln |B_L| - d \ln |B_R|,\\
 		  & = & \tau - d \lam - 2 d \ln |B_R| = \tau - d \lam. \label{tau_eta}
\eearr

All the Sachs data are determined by the $v$ data.\footnote{
The sign $\sg_L\sg_R$ is also needed to determine the Sachs data. This sign
is implicit in the specification of the $v$ data. The datum $e^{pq}$ on
each generator is given for a range of $v$ from $1$ to $\bar{v}$, and
$\sg_A$ is the sign of $\bar{v}^A - 1$.}
Note that even if it is not assumed
that the $v$ data matches a solution, the function $\eta(v)$, and thus the Sachs datum
of $e^{pq}(\eta)$, may be calculated from any regular $v$ data using (\ref{int_focusing1}).
Similarly, the remaining Sachs data may be obtained, from (\ref{drho_R}), (\ref{drho_L}),
(\ref{tau_eta}).

The transformation from regular $v$ data to Sachs data we have found is invertible.
Solving (\ref{drho_L}) and (\ref{tau_eta}) yields:
\bearr
\lam & = & \ln |\di_{\eta^R} \rho|_{S_0}| + \ln |\di_{\eta^L} \rho|_{S_0}|- 2\ln (2\rho_0) \\
\tau & = & \tau_\eta + d \ln |\di_{\eta^L} \rho|_{S_0}| - d \ln |\di_{\eta^R} \rho|_{S_0}|.
\eearr
(For Sachs data corresponding to regular $v$ data the derivatives of $\rho$ appearing as
denominators or arguments of logarithms do not vanish.)\footnote{
Sachs data obtained by setting
$B_R = 1$ satisfies $\di_{\eta^R} \rho|_{S_0} = 2\rho_0$. In general
$B_R = \frac{2\rho_0}{\di_{\eta^R} \rho|_{S_0}}$.}
The focusing equation
(\ref{focusing}) can be rewritten in the form
\be	\label{focusing2}
\di_\eta \di_\eta v = \frac{v}{8} \di_\eta e_{pq} \di_\eta e^{pq}.
\ee
Since $v^A = 1$ on $S_0$, and the Sachs data determine
$\di_{\eta^A} v^A|_{S_0} = \frac{1}{2\rho_0}\di_{\eta^A} \rho|_{S_0}$
and $e_{pq}(\eta^A)$, (\ref{focusing2}) has a unique solution $v^A(\eta^A)$ on each branch.
$v^A(\eta^A)$ and $e_{pq}(\eta^A, \theta)$ then determine the $v$ datum $e_{pq}(v^A,\theta)$,
showing that all $v$ data can be reconstructed from the Sachs data. If the Sachs data
was obtained by transforming regular $v$ data this inverse transformation yields the original
$v$ data. But the transformation relates the Sachs data and the $v$ data of a solution.
Thus if a solution matches the Sachs corresponding to a set of regular $v$ data then this
solution must also match the original $v$ data. This completes the demonstration of
the equivalence of regular $v$ data and Sachs data.


Let us turn to the diffeomorphism data.
The diffeomorphism data that will be used are $\bar{v}_A(\theta)$, the
area parameter at the endpoint on $S_A$ of the generator specified by
$\theta$, and $s_A^k = y_A^k(\theta)$, a map which gives the
position of this endpoint in a {\em fixed} chart $y_A$ on $S_A$.\footnote{
Spacetime is modeled by a manifold, and manifolds consist of individual {\em a priori}
identifiable points. It makes sense to compare different metrics at the same point (as
is done, for example, when varying the action), and one may distinguish between charts
that depend on the metric field, such as normal coordinates or our $v\theta$ chart,
and fixed charts which are, so to speak, painted on the manifold.
See appendix \ref{variations} for a more detailed discussion.}
The status of the $\bar{v}^A$ as a datum is curious. It is implicit in the
specification of the $v$ data on $\cN_A$, since it defines the range of
$v^A$ on which $e^{qp}$ is given, but it is not {\em functions} of the $v$
data. It is independent of the $v$ data if $e^{qp}$ is specified on a fixed,
reference range of $v\theta$ coordinates and $\bar{v}^A$ delimits the subset
of this range that corresponds to points on $\cN$. The entire set of data, consisting of
the diffeomorphism data $\bar{v}_A$ and $s_A$ and the $v$ data, is then free,
since both diffeomorphism data can be varied independently of the $v$ data by
acting on the spacetime metric with suitable diffeomorphisms, which of course
map solutions to solutions.

These data and their variations suffice to determine $\omega_\cN[\dg_1, \dg_2]$, 
when the variations $\dg_1$ and $\dg_2$ are what we will call ``admissible''.
(This is explained in detail in subsection \ref{expressing_with_null}.)
This is enough for our purposes because the evaluation of the Poisson
brackets of the data carried out in \cite{MR} and \cite{PRL} requiers the
the symplectic 2-form only on admissible variations.


The diffeomorphism data play a role in the calculation of the Poisson bracket
in \cite{PRL}, but are they essential? Should they be regarded as "physical"?
It seems to depend on what one wants to do. Because the observables defined in
the introduction are diffeomorphism invariant they do not depend on the
diffeomorphism data.\footnote{
Not all diffeomorphisms are gauge, that is, not all are generated by degeneracy 
vectors of the symplectic 2-form. Nevertheless, within any region which
excludes a neighborhood of the boundary $\di\cN$, any diffeomorphism may be 
realized as the restriction of a diffeomorphism which vanishes in a neighborhood
of $\di\cN$, and such diffeomorphisms are gauge, as will be shown in subsection 
\ref{diffeos}. Thus on the {\em interior} of the domain of dependence $D[\cN]$ of 
$\cN$ all gauge invariant degrees of freedom of the metric are diffeomorphism 
invariant. The observables seem to completely capture these degrees of freedom,
at least for metrics without isometries \cite{MR}.}
The requierment that the Poisson bracket of the data reproduce the brackets 
between observables, which (\ref{auxbracketdef}) ensures, can thus at most 
determine the brackets between the $v$ data. In subsection \ref{role_of_diff_data}
it will be shown directly that (\ref{auxbracketdef}) does not determine the brackets
of the diffeomorphism data, indeed it does not involve them at all.

It seems therefore that in the canonical theory of the gravitational field in the
domain of dependence of $\cN$ the diffeomorphism data have only an auxiliary role.
Indeed, the $v$ data are found to form a closed Poisson subalgebra, that is, their
brackets are functions only of $v$ data \cite{PRL}, so the diffeomorphism
data could be eliminated altogether from the canonical formalism.

On the other hand the diffeomorphism data may be relevant to quasi-local energy 
or other quantities associated with the boundary $\di\cN$. Note that the 
diffeomorphism data, unlike the remaining data, "know" about the boundary 
$\di\cN$. In the present work the diffeomorphism data are included in the initial
data because the expression for the symplectic 2-form used in the calculation of
the Poisson brackets in \cite{PRL} does depend on them, and it is the main aim of
the present work to present a derivation of this expression.

Before closing this subsection let us state precisely the complete set of data
to be used: It consists of
\begin{itemize}

\item 10 real $C^\infty$ functions, $\rho_0$, 
$\lam$, $\tau_p$, $\bar{v}_A$, and $s_A^i$, on a domain $D \in \mathbb{R}^2$ 
having the topology of a closed disk, with $\bar{v}_A > 0$ and $\neq 1$ 

\item two $C^\infty$, real, symmetric, unimodular $2 \times 2$ matrix 
valued functions ($e_{pq}$ on $\cN_L$ and $\cN_R$) on the domains 
$\{\theta \in D, \min(1,\bar{v}_A(\theta)) \leq v^A \leq \max(1,\bar{v}_A(\theta))\}$, 
$A = L,R$ which match at $v^L = v^R = 1$ (i.e. on $S_0$). 

\end{itemize}
Our phase space is the space of valuations of these data.

\section{The symplectic 2-form of the Einstein-Hilbert action}\label{symp1}

In the present section we will define the symplectic 2-form $\omega_\Sg$
of any oriented hypersurface $\Sg$ embedded in spacetime, and calculate it in
terms of the spacetime metric and its variations. In subsecuent sections this
expression is reduced to one in terms of our free initial data in the special
case that the hypersurface is an embedded double null sheet.

$\omega_\Sg$ will be defined for metrics $g$ that satisfy the vacuum field 
equations and variations that lie in the space $L_g$ of solutions to the field
equations linearized about $g$. Although $\omega_\Sg$ is an on shell quantity
it depends on the off shell action (on solutions the Einstein-Hilbert action is
zero!), and it is most naturally defined as a pullback to the space of 
solutions of a symplectic 2-form\footnote{
Recall that in the present work the term symplectic 2-form subsumes degenerate forms,
which are often called ``presymplectic'' in the literature.}
$\Omega_\Sg$ defined by the action
functional on all smooth metrics and variations. More precisely, $\omega_\Sg$
is the restriction of $\Omega_\Sg$ to metrics that satisfy the field
equations and to variations in $L_g$.\footnote{
At linearization stable solutions this is the pullback to the space of 
solutions since there $L_g$ coincides with the tangent space to the solution 
manifold. At non-linearization stable solutions $L_g$ is larger than the 
tangent space. According to the local linearizations stability theorem of 
\cite{Brill} all solutions are linearization stable in the interior of the 
domain of dependence of $\cN$. In fact, whether or not $L_g$ coincides with the
tangent space to the manifold of solutions will not affect our considerations.}
(See \cite{Lee_Wald} for the uses of $\Omega_\Sg$.)
 
The symplectic 2-form will be calculated from the Einstein-Hilbert action,
	\be		\label{EH}
                I = \frac{1}{16\pi G}\int_Q R \veg,
        \ee
where $\veg$ is the metric 4-volume form and $Q$ is the domain of
integration, which may be chosen freely.
The sign conventions for the curvature tensor and scalar are those of 
\cite{Wald}, that is, $R = R_{ab}{}^{ab}$ with
\be	\label{R_def}
[\nabla_a,\nabla_b]\bg_c = R_{abc}{}^d \bg_d
\ee
for any 1-form $\bg$.

The variation of the action due to a variation $\dg$ of the metric consists
of a bulk term, which vanishes on solutions, and a boundary term which 
determines the symplectic 2-form. The variation of ($16\pi G$ times) the
Einstein-Hilbert Lagrangean is
\be
\dg[R\veg] = [R_{ab} - \frac{1}{2} R g_{ab}]\dg g^{ab}\veg 
		+ \dg R_{ab}g^{ab}\veg,
\ee
where $R_{ab} = R_{acb}{}^c$ is the Ricci tensor. Clearly the first 
term vanishes on solutions. The second term is a divergence: From the
definition (\ref{R_def}) it follows that 
$\dg R_{abc}{}^d = -2\nabla_{[a} \dg \Cg^d_{b]c}$ so
\be
\dg R_{ab}g^{ab}\veg 
= -2\nabla_{a} \dg \Cg^{[c}_{cb}g^{a]b}\veg.
\ee
The integral of this divergence is the boundary term in the variation of the 
action. For any vector field $v$
\be
\nabla_a v^a \veg = d \wedge v \cont\veg,
\ee
so $\dg R_{ab}g^{ab}\veg = d\wedge \ag$ with\footnote{
We will occasionally mix abstract index notation with index free notation 
for differential forms. In particular abstract index notation will be used to 
indicate contractions of tensors. To avoid confusion when some indices 
of a tensor are written and other, uncontracted, indices are not, the unwritten 
indices are indicated by dots.}
\be 	\label{alpha_def}
\ag = -2 \dg \Cg^{[c}_{cb}g^{a]b}\veg_{a\cdot\cdot\cdot}
\ee

The boundary term in the variation of the action is thus
\be
    B[\dg] =  -\frac{1}{8\pi G}\int_{\di Q} \dg\Gamma^{[c}_{cb} g^{a]b}
                 \:\varepsilon_{a\cdot\cdot\cdot}, \label{boundaryEH}
\ee
The {\em symplectic potential} associated with a portion $\Sg$ of $\di Q$ is 
obtained by restricting the boundary integral (\ref{boundaryEH}) to $\Sg$: 
\be	\label{symppotential}
    \Theta_\Sg [\dg] =  -\frac{1}{8\pi G}\int_{\Sg} \dg\Gamma^{[c}_{c b} g^{a]b}
                         \:\varepsilon_{a\cdot\cdot\cdot}. 
\ee
The {\em symplectic 2-form} on a pair of variations $\dg_1$ and $\dg_2$ is
\bearr
\Omega_{\Sg}[\dg_1,\dg_2] & \equiv & \dg_1\Theta_{\Sg}[\dg_2] - \dg_2\Theta_{\Sg}[\dg_1]
                                - \Theta_{\Sg}[[\dg_1,\dg_2]] \label{Omega_def}\\
			  & = & -\frac{1}{8\pi G} \int_\Sg \delta_2
\Gamma^{[c}_{c b} \delta_1 (g^{a]b} \varepsilon_{a\cdot\cdot\cdot}) - (1 \leftrightarrow 2)\nonumber
\\
&& \label{presymp0}	
\eearr
See \cite{Crnkovic_Witten} and \cite{Lee_Wald}. $\Omega_{\Sg}[\dg_1,\dg_2]$ may be
interpreted as the curl of $\Theta_{\Sg}$ in the space of metric fields, 
evaluated on two tangent vectors, $\dg_1$ and $\dg_2$, to this space. 

The definition of $\Theta_\Sg$ given is in fact ambiguous. The boundary integral 
$B$ in the variation is quite unambiguously defined, but the integrand of
$B$ is not. Adding an exact form to it would not affect $B$, but would
alter $\Theta_\Sg$ by an integral over $\di\Sg$. There is also the freedom to 
add a boundary term to the action. At first sight it would seem that such a
boundary term only adds a total variation to $\Theta_\Sg$, which would not 
affect $\Omega_\Sg$. 
However whether this is so actually depends on the precise prescription used to
determine the integrand of $\Theta_\Sg$ from the Lagrangean. Lee and Wald
\cite{Lee_Wald} give such a prescription (in which boundary terms added to the
action {\em can} produce boundary terms in $\omega_\Sg$ if they depend on
derivatives of the fields).
Our expression (\ref{symppotential}) for $\Theta_\Sg$ corresponds to the 
Einstein-Hilbert action without boundary term according to this prescription. 
But is there a physical reason to prefer the Lee-Wald prescription? Are boundary 
terms in $\omega_\Sg$ important? 

The Poisson bracket should not depend on boundary terms. The Peierls
bracket is expressed directly in terms of the advanced and retarded Green's
functions, which are not affected by boundary terms in the action.\footnote{
The Greens functions depend only on the field equations derived from the action
with a suitable source term. The boundary terms we are considering are ones like
the York-Gibbons-Hawking term, which are matched to boundary conditions on the
variations of the fields so that the presence of the boundary does not affect the
field equations that result from extremizing the action. We are not considering
boundary terms which represent a physical feature at the boundary, and of course
would affect Greens functions.}
The Poisson bracket $\{\cdot,\cdot\}_\bullet$ on initial data calculated in
\cite{MR} and \cite{PRL}, which is defined by the requierment that it reproduce
the Peierls bracket, should also be insensitive to boundary terms.
Indeed the condition (\ref{auxbracketdef}) which ensures the matching to the Peierls
bracket is manifestly unaffected by the addition of boundary terms to the symplectic
2-form.\footnote{
The condition (\ref{auxbracketdef}) does not determine the brackets of all the data uniquely.
In \cite{PRL} the bracket is therefore derived from a strengthened version of (\ref{auxbracketdef})
which could be affected by boundary terms in the action. But as long as no compelling motivation
is found for the auxiliary conditions used to obtain a unique bracket, any such sensitivity to
boundary terms has to be regarded as artificial.}
Note that brackets obtained in \cite{PRL} do not ``know'' where the boundary $\di\cN$ is,
That is, they are unchanged by a displacement of the boundary, except in the case of the
brackets of the diffeomorphism data which themselves encode features of the boundary.

On the other hand, the canonical generators of diffeomorphisms that move the 
boundary $\di\Sg$ {\em do} seem to depend on boundary terms in $\omega_\Sg$. 
Such generators define quasi-local notions of energy, angular momentum, etc. and
the correct boundary terms would presumably be defined by the properties one 
wants these quasi-local quantities to have. This interesting direction will not 
be explored here. Rather we shall simply adopt the symplectic potential 
(\ref{symppotential}) corresponding to the Einstein-Hilbert action without
boundary term. 

\subsection{diffeomorphisms}\label{diffeos}

The degeneracy vectors of the symplectic 2-form are variations $\Dg$
such that $\omega_\Sg[\Dg,\dg] = 0$ for all smooth solutions to the linearized
field equations $\dg$. These are often called {\em gauge variations} although 
it is not clear that this is the most apropriate definition of ``gauge" when 
$\Sg$ has boundaries. In general relativity the degeneracy variations of the 
metric are Lie derivatives of the metric along vector fields satisfying certain 
conditions at $\di\Sg$.
This is the familiar diffeomorphism invariance of 
general relativity: If $\psi_t$ is a family of diffeomorphisms parameterized 
by $t \in \Real$ then the $t$ derivative of the image metric $\psi^*_t(g)$ is 
$d\psi^*_t(g)/dt = -\pounds_v g$, where $v$ is the field of tangents to the 
orbits of the manifold points under $\psi_t$, so Lie derivatives generate 
diffeomorphisms. 

Let us evaluate $\omega_\Sg[{\pounds_v},\dg]$ for any $C^\infty$ vector 
field $v$ on spacetime and $\dg \in L_g$. 
\be	\label{omega_diffeo}
	\omega_\Sg[{\pounds_v},\dg] = {\pounds_v} \Theta_\Sg[\dg] 
	- \dg \Theta_\Sg[{\pounds_v}] - \Theta_\Sg[[\pounds_v,\dg]].
\ee
Now 
\be
	\Theta_\Sg[\dg]  =  \frac{1}{16\pi G}\int_{\Sg} \ag
\ee
with $\ag$ the 3-form defined in (\ref{alpha_def}). Thus
\bearr
	{\pounds_v} \Theta_\Sg[\dg] & = &\frac{1}{16\pi G}\int_\Sg {\pounds_v} \ag\\
				    & = &\frac{1}{16\pi G}\int_\Sg v\cont 
	[d\wedge \ag] + d \wedge [v\cont\ag].
\eearr
But $d \wedge \ag = \dg R_{ab}g^{ab}\veg$ is the divergence term 
in the variation of the Einstein-Hilbert Lagrangean density, which vanishes 
because $\dg$ satisfies the linearized vacuum field equation $\dg  R_{ab} = 0$. 
Therefore
\bearr
	\lefteqn{{\pounds_v} \Theta_\Sg[\dg]}\\
 	& = & \frac{1}{16\pi G}\int_{\di\Sg} v\cont\ag \\
	& = & -\frac{1}{8\pi G}\int_{\di\Sg} v^b 
	\dg\Gamma^{[c}_{cd} g^{a]d} \:\veg_{ab\cdot\cdot} \\
	& = & \frac{1}{8\pi G}\int_{\di\Sg} v^a 
	(\nabla^b \dg \veg_{ab\cdot\cdot} + \frac{1}{2}\nabla_c \dg g^{bc}\veg_{ab\cdot\cdot}).
\label{L_v_Theta_delta}
	\eearr						
In the last line the identity 
\be \label{dGamma_id}
\dg\Cg^c_{ab} = \frac{1}{2}g^{cd}\{\nabla_b \dg g_{da} + \nabla_a \dg g_{db} 
- \nabla_d \dg g_{ab}\}.
\ee
has been used.  

The second term in (\ref{omega_diffeo}) is the $\dg$ variation of
	\be
		\Theta_\Sg[{\pounds_v}] = -\frac{1}{8\pi G}
		\int_{\Sg} {\pounds_v}\Gamma^{[c}_{cb} g^{a]b}\:\varepsilon_{a\cdot\cdot\cdot}
 	\ee
But (\ref{dGamma_id}) and Einstein's field equation, which $g$ satisfies, 
imply that
\bearr
	\lefteqn{{\pounds_v}\Gamma^{[c}_{cb} 
		g^{a]b}\veg_{a\cdot\cdot\cdot}} \nonumber\\
	& = & (\nabla^a \nabla_c v^c -\frac{1}{2}\nabla_c \nabla^c v^a 
		-\frac{1}{2}\nabla_c \nabla^a v^c)\veg_{a\cdot\cdot\cdot} \\
	& = & \nabla_c \nabla^{[a} v^{c]}\veg_{a\cdot\cdot\cdot}	\\
	& = & \frac{1}{2} d \wedge (\nabla^a v^b \veg_{ab\cdot\cdot}).
\eearr
Thus
\be		\label{diffTheta1}
		\Theta_\Sg[{\pounds_v}] = -\frac{1}{16\pi G} \int_{\di\Sg} 
			\nabla^a v^b\:\veg_{ab\cdot\cdot} 
\ee
Since
\be
	[\pounds_v,\dg]g = -\pounds_{\dg v}g
\ee
it follows that
\be	\label{commutator_term}	
	\Theta_\Sg[[\pounds_v,\dg]] = \frac{1}{16\pi G}
		\int_{\di\Sg} \nabla^a \dg v^b\:\veg_{ab\cdot\cdot}. 
\ee

Subtracting (\ref{commutator_term}) and the $\dg$ variation of (\ref{diffTheta1})
from (\ref{L_v_Theta_delta}) one obtains
\bearr
\lefteqn{\omega_\Sg[\pounds_v,\dg]} \nonumber\\ 
& = & \frac{1}{16\pi G} \int_{\di\Sg} v^a(2\nabla^b \dg\veg_{ab\cdot\cdot}
      + \nabla_c \dg g^{bc}\veg_{ab\cdot\cdot}) \nonumber\\
&& \ \ \ \ \ \ \ \ \ \ \  + \dg (\nabla^a v^b \veg_{ab\cdot\cdot}) 
	- \nabla^a \dg v^b \veg_{ab\cdot\cdot},  \label{boundary_diff2}	\\
& = & \frac{1}{16\pi G} \int_{\di\Sg} 3 v^{[a}\dg\Gamma^c_{cd} g^{b]d}
	\veg_{ab\cdot\cdot}
      + \dg[g^{ca}\veg_{ab\cdot\cdot}]\nabla_c v^b.    \label{boundary_diff}
\eearr

This integral obviously vanishes when $v$ and $\nabla v$ vanish on $\di\Sg$. The
corresponding variation ${\pounds}_v$ is therefore a degeneracy vector of the
symplectic 2-form.

\section{The symplectic 2-form on $\cN$ in terms of the free null data.}\label{symp_null}

In this section the symplectic 2-form $\omega_\cN[\dg_1,\dg_2]$ defined in
section~\ref{symp1} will be expressed in terms of the free null initial data
defined in subsection~\ref{data}, for variations $\dg_1$ and $\dg_2$ that
satisfy the linearized field equations and a series of further conditions that 
define what we will call the ``admissible variations''.

Admissible variations are fairly special, but our expression for $\omega_\cN[\dg_1,\dg_2]$
will in fact be applicable to a much larger class of variations. We will show that any
pair of variations, $\dg_1 \in L_g$ and $\dg_2 \in L_g^0$, may be replaced in $\omega_\cN[\dg_1,\dg_2]$
by corresponding admissible variations without changing the value of the symplectic
2-form. Our expression for the symplectic 2-form in terms of the free null data therefore
suffices to convert (\ref{auxbracketdef}) into an explicit condition on the Poisson brackets
of these data.

In \cite{PRL} the Poisson brackets of the initial data are obtained from a somewhat strengthened 
version of (\ref{auxbracketdef}), which can also be expressed in terms of the initial data
using the expression for $\omega_\cN[\dg_1,\dg_2]$ on admissible variations.  

The first subsection treats conceptual issues involved in expressing $\omega_\cN[\dg_1,\dg_2]$
in terms of the null data of \ref{data} and demonstrates that attention may be restricted to the
class of admissible variations. The next subsection demonstrates the limited role of the
diffeomorphism data defined in \ref{data}.
The third subsection presents some charts used in the calculations. In the fourth subsection
the symplectic potential is evaluated in terms of the free null data. Finally, in the last
subsection, this expression for the symplectic potential is used to calculate
the symplectic 2-form in terms of the free null data.

\subsection{Variations in terms of null initial data and admissible variations}\label{expressing_with_null}

According to (\ref{presymp0}) the symplectic 2-form on $\cN$, at a given spacetime metric
$g$, is
\be	\label{presymp0b}
\omega_{\cN}[\dg_1,\dg_2] = -\frac{1}{8\pi G} \int_\cN \delta_2
\Gamma^{[c}_{c b} \delta_1 (g^{a]b} \varepsilon_{a\cdot\cdot\cdot}) - (1 \leftrightarrow 2),
\ee
where $\dg_1 g$ and $\dg_2 g$ are solutions to the field equations linearized about $g$.
Our task is to express $\omega_{\cN}[\dg_1,\dg_2]$ in terms of the free null initial data
and their variations in the case that $\cN$ is a double null sheet of $g$, $\dg_1 \in L_g$,
and $\dg_2 \in L_g^0$, the set of solutions to the linearized field equations that vanish
in a spacetime neighborhood of $\di\cN$.

It is not {\em a priori} obvious that this can be done. By definition 
the spacetime metrics matching the null data make the hypersurfaces
$\cN_L$ and $\cN_R$ null, so the variations of these data only parametrize
variations $\dg \in L_g$ that preserve the nullness of $\cN_L$ and $\cN_R$.
\footnote{
It {\em is} possible to define variations of null data under general
variations of the metric, if the null data live not on $\cN$ but on a metric
dependent double null sheet associated with $\cN$. Working along these
lines one arrives ultimately at the same theory presented here. }
Arbitrary variations will not in general do this for a given, fixed, hypersurface
$\cN$. That the symplectic 2-form $\omega_{\cN}[\dg_1,\dg_2]$ can nevertheless be
expressed in terms of the variations of null data for all $\dg_1 \in L_g$ and
$\dg_2 \in L_g^0$ is a consequence of the diffeomorphism gauge invariance of general
relativity.

Although the branches of the fixed hypersurface $\cN$ may cease to be null when the spacetime
metric is changed slightly, it is always possible, by a small deformation
of $\cN$, to obtain a new hypersurface $\cN'$ which is a double null
sheet of the new metric. (The double null sheet $\tilde\cN_{S_0}$ swept
out by the future null normal geodesics from $S_0$ in the new metric is an example.)
Thus, if the given change in the metric is followed by the action on the metric of a suitable
diffeomorphism, which moves $\cN'$ to $\cN$, then the resulting total alteration of the metric
preserves the double null sheet character of $\cN$. Any variation $\dg$ may therefore
be split into the sum of a null sheet preserving variation $\dg'$, that is, one that
preserves the null sheet character of $\cN$, and a diffeomorphism generator $\pounds_u$.

Applying this decomposition to the two arguments $\dg_1, \dg_2 \in L_g$ of the
symplectic 2-form one obtains
\be	\label{omega_decomp}
\omega_\cN[\dg_1, \dg_2] = \omega_\cN[\dg'_1, \dg'_2]
+ \omega_\cN[\dg'_1, \pounds_{u_2}]
+ \omega_\cN[\pounds_{u_1}, \dg'_2]
+ \omega_\cN[\pounds_{u_1}, \pounds_{u_2}].
\ee
If the diffeomorphism generators are degeneracy vectors of $\omega_\cN$,
that is, if they are gauge, then all terms but the first vanish, and in
this first term only the null sheet preserving variations $\dg'$ appear.
In this case, the fact that the variations of null data can only
parametrize nullness preserving variations would not be an impediment to
expressing the symplectic 2-form in terms of these data. Indeed, the $v$ data
of subsection \ref{data} determines the metric and its first derivatives on
$\cN$ up to diffeomorphisms that map
$\cN$ to itself. If the generators of all such diffeomorphisms were
degeneracy vectors then the $v$ data and their variations would suffice by
themselves to determine $\omega_\cN[\dg'_1, \dg'_2]$; The $v$ data and their
variations would determine the metric and its derivatives, and the gauge
equivalence class of their variations under $\dg'_1$ and $\dg'_2$, up to a
diffeomorphism mapping $\cN$ to itself, and the integral (\ref{presymp0b}) is
invariant under such diffeomorphisms.

However, not all diffeomorphism generators are degeneracy vectors of the
symplectic 2-form. Eq (\ref{boundary_diff}) shows that the diffeomorphism
terms in (\ref{omega_decomp}) are integrals over the boundary of $\cN$ that
might not vanish. Indeed some $\dg \in L_g$ might not be gauge equivalent to
{\em any} null sheet preserving variation. That is, there might exist no null sheet
preserving variation $\dg'$ such that $\dg' - \dg$ is a degeneracy vector.
(In fact it seems plausible that this is the case for some $\dg$, but it has
not been demonstrated.) If this is so then $\omega_\cN[\dg_1, \dg_2]$ cannot be
expressed in terms of null initial data for all $\dg_1, \dg_2 \in L_g$.

Fortunately we do not need to express $\omega_\cN[\dg_1, \dg_2]$ in terms of null
initial data for completely general $\dg_1$ and $\dg_2$ in $L_g$. We are interested 
in the case in which $\dg_1$ is arbitary but $\dg_2 g_{ab}$ vanishes
in a spacetime neighborhood of $\di\cN$,\footnote{
It seems that the same results can be obtained with the weaker condition that 
$\dg_2 g_{ab}$ and $\dg_2 \nabla_c g_{ab}$ vanish on $\di\cN$ itself. We will 
not pursue this matter here.}
that is, $\dg_1 \in L_g$, $\dg_2 \in L_g^0$, because this is the case relevant for 
the calculation of the Poisson bracket via (\ref{auxbracketdef}) in \cite{PRL}.

Let us suppose then that $\dg_2 \in L_g^0$. This restriction implies that the
diffeomorphism terms in (\ref{omega_decomp}) do vanish: If $\dg_2$ vanishes in a 
neighborhood of $\di\cN$ then $\omega_\cN[\pounds_{u_1},\dg_2]$ is zero 
because it is an integral over $\di\cN$ of an integrand proportional to
$\dg_2 g_{ab}$ and its derivatives there. Furthermore, when $\dg_2$ vanishes in a
neighborhhood of $\di\cN$ the field $u_2$ may be chosen so that it also vanishes
in a (generally different) neighborhood of $\di\cN$ (see below). This implies that
$\omega_\cN[\dg'_1,\pounds_{u_2}]$ is also zero. Thus, when $\dg_2 \in L_g^0$
\be	\label{omega_null}
\omega_\cN[\dg_1, \dg_2] = \omega_\cN[\dg'_1, \dg'_2]. 
\ee
This means that $\omega_\cN[\dg_1, \dg_2]$ can be expressed in terms of null
initial data, that is, in terms of data sufficient to determine the metric 
and its derivatives up to gauge on $\cN$ assuming $\cN$ is a double null sheet. 
In section \ref{symp2} such an expression is given explicitly, in terms of the 
free null data defined in subsection \ref{data}.

Before continuing let us return to the diffeomorphism generator 
$\pounds_{u_2} = \dg_2 - \dg_2'$ and show that $u_2$ may indeed be chosen 
so that it vanishes in a neighborhood of $\di\cN$. 
To this end we define a new metric dependent double null sheet $\tilde\cN_{\di\cN}$
swept out by past normal null geodesics from $S_L$ and $S_R$ rather than future normal
null geodesics from $S_0$: The generators of $\cN$ may be regarded as normal null geodesics
emerging to the past from $S_L$ or $S_R$ and truncated where they meet at $S_0$. When the
metric is changed these past normal null geodesics from $S_L$ and $S_R$ are
also changed, and sweep out new null hypersurfaces $\tilde\cN_{S_L}$ and 
$\tilde\cN_{S_R}$. If $\dg$ is a variation that vanishes in a neighborhood $\cal W$ of 
$\di\cN$ then it will not disturb the geodesics that make up the portion
$\di\cN - S_L - S_R$ of the boundary of $\cN$, and these will still meet at $\di S_0$.
Furthermore, if the change in the metric is small enough $\tilde\cN_{S_L}$ and $\tilde\cN_{S_R}$ will
intersect on a disk, $\tilde S_0$, where they may be truncated, and thus truncated will
contain no caustics. Thus $\tilde\cN_{\di\cN} = \tilde\cN_{S_L} \cup \tilde\cN_{S_R}$ is a double null sheet
of the perturbed metric. It is clear that the perturbed and unperturbed generators 
from $S_A$ coincide until they leave $\cal W$. Thus $\tilde\cN_{\di\cN}$ coincides with $\cN$ in a
neighborhood of $S_A$, and also in a neighborhood of $\di\cN - S_L -S_R$, since 
generators sufficiently near $\di\cN - S_L -S_R$ never leave $\cal W$. (This follows from
the compactness of the generator segments that sweep out $\di\cN - S_L -S_R$.)\footnote{\label{inv_gen_footnote}
The branch $\cN_A$ of $\cN$ is the image under the exponential map of a compact solid
cylinder $N_A$ in the normal bundle of $S_A$, the generators being the images of
parallel straight null lines in $N_A$ which will also be called generators.
The preimage $Z \subset N_A$ of the subset $\cal W\cap \cN_A$ of $\cN_A$ on which the
metric is invariant is open in $N_A$, since $\cal W\cap \cN_A$ is open in $\cN_A$ and
the exponential map is continuous. (Here the open sets in a subset $S$ of an ambient
space $X$ are the intersections of open sets of $X$ space with the subset $S$.)
Thus $Z$ can be expressed as a union of open solid cylinders of the form $c = l \times x$,
with $l$ an open line segment parallel to the generators and $x \subset S_A$ is in $S_A$.
Since any generator from $\di S_A$ lies in $Z$ it is covered by these cylinders.
But since it is compact it has a finite subcover ${c_i}$. The intersection $y = \cap_i x_i$
is a neighborhood of the base point of the generator in $\di S_A$, open in $S_A$, such that
generators from $y$ lie entirely in $Z$. Taking the union of such $y$s one obtains an
open neighborhood $Y$ of $\di S_A$ in $S_A$ such that all generators from $Y$ remain in
$Z$ until they leave $N_A$. See \cite{MR} proposition B.8. for a different proof.}
As a consequence $\tilde\cN_{\di\cN}$ can be mapped to $\cN$ by a diffeomorphism that
reduces to the identity in a neighborhood of $\di\cN$. That is, $\dg_2 \in L_g^0$ implies
that $u_2$ may be chosen to vanish in such a neighborhood. 

Our fundamental condition defining the Poisson bracket (\ref{auxbracketdef}), 
$\dg A = \omega_\cN[\{A,\cdot\}_\bullet, \dg]\ \ \forall \dg \in L_g^0$,
may also be expressed in terms of the null initial data.
The variation $\{A,\cdot\}_\bullet$ is already null sheet preserving by virtue
of its definition: The bracket $\{\cdot,\cdot\}_\bullet$ is a Poisson bracket
on the null initial data, so $\{A,\cdot\}_\bullet$ is a variation of these
initial data, which of course defines a null sheet preserving variation of
the spacetime metric (up to diffeomorphisms which do not affect the value of
$\omega_\cN[\{A,\cdot\}_\bullet,\dg]$ when $\dg \in L_g^0$).
(See \cite{MR} appendix C.) The variations
$\dg$ may be restricted to null sheet preserving variations without weakening
the condition on $\{A,\cdot\}_\bullet$ that (\ref{auxbracketdef}) implies:
The variation $\dg$ may be replaced by $\dg'$ on both sides of the equation
without altering the value of either, on the left because the observable $A$
is diffeomorphism invariant, and on the right because of (\ref{omega_null}).
Finally, any variation of $A$ may be written as a sum of the corresponding
variations of the initial data integrated against suitable smearing functions.
The smearing functions are the functional derivatives of $A$ by the initial data,
which are well defined because $A$ is functionally differentiable in the spacetime
metric, and variations of the metric satisfying the linearized field equations
are determined, up to diffeomorphisms, by the variations of the initial data.
(See \cite{MR} appendix C.)
$A$ may thus be replaced in (\ref{auxbracketdef}) by a sum of smeared null
initial data, yielding an equation entirely in terms of the variation $\dg'$
of the null initial data, and the Poisson brackets of these data.\footnote{
As mentioned in a previous footnote, an alternative point of view is possible, in which
the variations are not restricted to be null sheet preserving, but rather the definitions
of the null data are extended to geometries in which $\cN$ is not null. We will not adopt
this point of view but let us sketch it here: Suppose $\dg$ is a, not necessarily null sheet
preserving, variation. Recall that the action $\dg'\varphi(\theta,v)$ of a null sheet
preserving component $\dg' = \dg - \pounds_u$ of $\dg$ on the a null datum $\varphi(\theta,v)$
on $\cN$ is equal to action of $\dg$ on the same datum on a double null sheet
$\cN'$ that varies with the metric.
A choice of this double null sheet suitable for $\dg_1$ is $\cN' = \cN_{S_0. g}$,
since it is defined for all variations in $L_g$. For $\dg_2$ a suitable choice is
$\cN' = \tilde\cN_{\di\cN}$, since it is defined for all variations in $L_g^0$ and
corresponds to $u_2 = 0$ in a neighborhood of $\di\cN$. With this interpretation of
the null data in the variations the explicit expression for the symplectic 2-form in
terms of these data obtained in section \ref{symp2} applies directly to any pair of
variations $\dg_1 \in L_g$, $\dg_2 \in L_g^0$, whether they are null sheet preserving or not.

Condition (\ref{auxbracketdef}) reduces to an equation on the Poisson
bracket on the null data on $\tilde\cN_{S_0}$ as follows: As we have seen,
$\omega_\cN[\{A,\cdot\}_\bullet, \dg]$ may be expressed in terms of
the $\dg$ variations of data on $\tilde\cN_{\di\cN}$ and the variations under
$\{A,\cdot\}_\bullet$ of data on $\tilde\cN_{S_0}$. Furthermore $\dg A$ may be
expressed as a sum of the $\dg$ variations of the data on $\tilde\cN_{\di\cN}$,
smeared with the functional derivatives of $A$ by these data. Now note that
the functional derivatives of $A$ by the data on $\tilde\cN_{\di\cN}$ and on
$\tilde\cN_{S_0}$ are in fact the same, because the variation of the metric
produced by a variation of the data on $\tilde\cN_{\di\cN}$ and that produced
by the same variation of the data on $\tilde\cN_{S_0}$ differ by a diffeomorphism,
and $A$ is diffeomorphism invariant. Thus $\{A,\cdot\}_\bullet$ can be expanded
into a sum of the Poisson actions of the initial data on $\tilde\cN_{S_0}$, smeared
with the same functions (of $\theta$ and $v$) as appear in the expansion of $\dg A$
in terms of variations of the data on $\tilde\cN_{\di\cN}$.}

The requierment that $\dg'_1 =  \dg_1 - \pounds_{u_1}$ preserves the double null
sheet character of $\cN$ leaves considerable freedom in the choice of $u_1$. This
freedom will be exploited to restrict the variations we have to consider still
further. We will require
\begin{itemize}
 \item[1] that the variations map the generators that lie in the boundary $\di\cN$
to themselves,
 \item[2] that they leave invariant the area density $\bar{\rho}$ in the fixed
chart $y_A$ on the truncation surface $S_A$ of each branch,
\end{itemize}
and finally,
\begin{itemize}
 \item[3] that they leave invariant a special chart constructed from the
metric field in a spacetime neighborhood of each truncation surface $S_A$.
\end{itemize}
These charts, the $a_L$ and $a_R$ charts defined in subsection \ref{charts}, will
play an important role in the evaluation of the symplectic 2-form in terms of null data.

All these conditions already hold for $\dg'_2$ because this variation leaves the entire
metric field invariant in a spacetime neighborhood of $\di\cN$. They can be made to hold for
$\dg'_1$ by adding a suitable diffeomorphism generator, that is, by adjusting $u_1$:
If $\dg'_1$ perturbs the generators in $\di\cN$ then clearly a suitable diffeomorphism
returns them to their unperturbed courses. If $\dg'_1$ alters the $y$ chart area density
$\rho_y = \det [\di\theta/\di y] \rho$ at the endpoint of a generator on $S_A$ then
the generator can always be extended or shortened so that $\rho_y$ at the new endpoint
equals the unperturbed value of $\rho_y$ at the old endpoint on $S_A$, because
$\rho_y \propto v^2$ is nowhere stationary along the generator.
The generators thus lengthened or shortened can then be mapped to the original generators
of $\cN_A$ by a diffeomorphism. 

It remains only to ensure that the $a_A$ chart is preserved
by $\dg'_1$ in a neighborhood of $S_A$. Clearly this can be done by adding a diffeomorphism
generator to $\dg'_1$. What has to be shown is that it can be done without violating the
other conditions on $\dg'_1$. Let us suppose then that $\dg'_1$ preserves the double null sheet
character of $\cN$, and that it satisfies conditions 1 and 2.

As will be explained in 
subsection \ref{charts} the $a_A$ chart is an extension to a spacetime region of a chart on 
$\cN_A$, formed from the coordinates $y^i_A$, $r = v/\bar{v}$, and a fourth coordinate, $u$.
The $y^i$ label the generators, with each generator taking the values of $y^i$ of its endpoint
on $S_A$, while $r$ labels the points within each generator. Finally, $u$ is a coordinate transverse
to the hypersurface swept out by the generators. It vanishes on the generators themselves.

On $S_A$ itself the coordinates $y^i_A$ are fixed by definition, and $\dg'_1$ preserves them
on $\di\cN_A - S_A - S_0$ because of condition 1. It also preserves $u = 0$ on $\cN_A$ because
it is null sheet preserving, implying that the generators remain in $\cN_A$.
Thus condition 3, that the variation leaves invariant all the $a$ coordinates in a spacetime 
neighborhood of $S_A$, can be realized by adding a diffeomorphism generator which leaves 
$\di\cN_A$ invariant and maps $\cN_A$ to itself (that is, one that corresponds to
a vector field that vanishes on $\di\cN_A$ and is tangent to $\cN_A$ on the remainder of
$\cN_A$). But such a diffeomorphism generator clearly preserves conditions 1 and 2, and the null 
sheet character of $\cN_A$.

The null sheet preserving variations satisfying conditions 1,2, and 3 will be called {\em admissible}
variations.\footnote{
In \cite{PRL} a somewhat smaller set of variations was termed ''admissible``.}
In subsection \ref{symp_pot} the symplectic potetial $\Theta[\dg]$ is calculated in terms of
our null data on admissible variations. Then, in subsection \ref{symp2}, the symplectic 2-form is
calculated from the symplectic potential via (\ref{Omega_def}), again for admissible variations.
This is possible because the commutator of admissible variations is also admissible.

\subsection{The limited role of the diffeomorphism data}\label{role_of_diff_data}

What is the role of the ``diffeomorphism data'' introduced in subsection \ref{data}? Recall
that the free data defined in subsection \ref{data} consists of the so called ``$v$ data'',
which are equivalent to Sachs' free null initial data, as well as the diffeomorphism data
$s_A$ and $\bar{v}_A$. These latter data constitute partial information about how the $v\theta$
charts, to which the $v$ data are referred, are placed on $\cN$.

The diffeomorphism data appear in the expression for the symplectic 2-form found in
\cite{MR} and used in \cite{PRL}. Indeed Poisson brackets are calculated for them.
However, it was also argued in subsection \ref{data} that the diffeomorphism data
are not essential to the canonical formulation of general relativity in the domain
of dependence of $\cN$. They do not affect the spacetime geometry in the domain
of dependence, nor the so called observables, which are functionals of the
geometry, nor the Poisson brackets between these observables. Thus the condition
(\ref{auxbracketdef}), which ensures that the Poisson bracket on the data reproduces the
Peierls brackets of the observables, ought not define brackets for the diffeomorphism data.

Here this expectation will be confirmed. It will be shown that the symplectic 2-form
$\omega_\cN[\dg_1,\dg_2]$ does not depend on the variations of the diffeomorphism data if
$\dg_2$ vanishes in a neighborhood of $\di\cN$, and that (\ref{auxbracketdef}) provides
no information about the Poisson brackets of the diffeomorphism data.

The diffeomorphism data will nevertheless be retained in the present work. This is done
mainly for consistency with \cite{PRL}, which the present work underpins. In \cite{PRL} a
strengthened version of (\ref{auxbracketdef}), in which the test variation $\dg$ need not
vanish in a neighborhood of $\di\cN$, is used to define a Poisson bracket on all the free
data of subsection \ref{data}, including the diffeomorphism data. This strengthened condition
requires an expression for $\omega_\cN[\dg_1,\dg_2]$ in terms of the null initial data valid for
{\em all} admissible variations. It is this expression, which depends on the variations of the
diffeomorphism data $s_A$, that is obtained in subsection \ref{symp2}.


Let us turn to the demonstration of the claims made above. Recall that when $\dg_2 \in L_g^0$
one may replace $\dg_1$ and $\dg_2$ in $\omega_\cN[\dg_1,\dg_2]$ by corresponding admissible
variations without changing the value of $\omega_\cN[\dg_1,\dg_2]$, and that the admissible
variation corresponding to $\dg_2$ still lies in $L_g^0$. Thus we may restrict
our attention to admissible $\dg_1$ and $\dg_2$ without loss of generality.

The variations of the $v$ data determine $\dg g_{ab}$ and $\nabla_c \dg g_{ab}$ on $\cN$ up to 
diffeomorphism generators. So they characterize $\dg_1$ sufficiently for the calculation of
the symplectic 2-form $\omega_\cN[\dg_1,\dg_2]$ when $\dg_1$ and $\dg_2$ are admissible variations
in $L_g$ and $L_g^0$ respectively. 
The situation is a little more subtle for $\dg_2$. Since $\omega_\cN[\dg_1,\dg_2]$ is not
invariant under the addition of non gauge diffeomorphism generators to $\dg_2$, the variation
under $\dg_2$ of non gauge diffeomorphism degrees of freedom must be specified. The diffeomorphism
data, $s_A$ and $\bar{v}_A$, measure such degrees of freedom. However, because $\dg_2 g_{ab}$ is
required to vanish in a spacetime neighborhood of $\di\cN$ the variations of the $v$ data under 
$\dg_2$ in fact determine those of the diffeomorphism data modulo gauge. Thus ultimately
$\omega_\cN[\dg_1,\dg_2]$ depends only on the variations under $\dg_1$ and $\dg_2$ of the $v$
data, and of course the unperturbed values of the $v$ data and of $\bar{v}_A$. (It does not depend
on the unperturbed values of $s_A$ because $s_A$ can be set to any desired value by a diffeomorphism
that maps $\cN$ to itself, and $\omega_\cN[\dg_1,\dg_2]$ is invariant under such diffeomorphisms.)

How does this come about? Because $\dg_2$ vanishes in a neighborhood of $\di\cN$ the $y$ chart
area density on $S_A$, $\bar{\rho}_A(y)$, is invariant under $\dg_2$.  Thus the variation of
\be
\bar{v}_A(\theta) = \sqrt{\frac{\bar{\rho}_A(s_A(\theta))\det[\di s_A/\di\theta]}{\rho_0}}
\ee
is determined by those of $\rho_0$ and $s_A$.\footnote{
In fact, by definition all admissible
variations leave $\bar{\rho}_A(y)$ invariant, so the variations of $\bar{v}_A$ can be eliminated
from $\omega_\cN[\dg_1,\dg_2]$ whenever $\dg_1$ and $\dg_2$ are admissible, even if neither lies
in $L_g^0$. Precisely this will be done in our calculation of the symplectic 2-form on admissible
variations.}

It remains to show that the variations of $s_A$ under $\dg_2 \in L_g^0$ are determined by those of the $v$ data.
In fact there is a trivial sense in which $s_A$ can vary independently of the $v$ data when
there is enough symmetry. The field $s_A$ depends on the choice of $\theta$ coordinates on $S_0$, which
is a gauge choice in our formalism. If the spacetime geometry near $\cN$ admits an isometry, a rotation,
that maps $\cN$ to itself, then it is possible to change $s_A$ without changing the $v$ data, by rotating
the $\theta$ chart. But of course such a variation is pure gauge. It does not contribute to the symplectic
2-form because it does not change the spacetime metric components or their derivatives at any point of $\cN$,
and the symplectic 2-form depends only on the variations of the spacetime metric. Such gauge variations will
be eliminated by holding the $\theta$ chart fixed in the variations we consider. As we will now see, once this
restriction is imposed $\dg_2 s_A$ is indeed determined by the corresponding variations of the $v$ data.

Suppose $\dg$ and $\tilde\dg$ are two admissible variations in $L_g^0$ that induce the same
variations of $v$ data. If the $v\theta$ charts on the braches of $\cN$ are given then 
the $v$ data determine the metric and its derivatives on each branch $\cN_A$ up to 
diffeomorphisms that fix the points of $\cN$. If the $v$ data are given but the placement
of the $v\theta$ charts is not specified, then there is of course an additional freedom 
in the metric corresponding to movements of this chart.
It follows that $\dg g_{ab} - \tilde\dg g_{ab} = \pounds_\xi g_{ab}$ where the vector field $\xi$
generates a diffeomorphism that maps $\cN$ to itself, and furthermore that on $\cN$ the field
$\xi$ reduces to the difference in the $\dg$ and $\tilde\dg$ variations of the $v$ and $\theta$
coordinates: $\dg \theta^p - \tilde\dg \theta^p = \xi^p$, $\dg v - \tilde\dg v = \xi^v$.

This has two immediate consequences. First, since the $\theta$ chart (and of course also $v = 1$) is
fixed on $S_0$, $\xi = 0$ there. Second, since $\dg - \tilde\dg \in L_g^0$, the diffeomorphism generated
by $\xi$ must reduce to an isometry in a neighborhood $U$ of $\di\cN$. But since $\xi$ vanishes on $S_0$,
$S_0 \cap U$ is fixed under the isometry, as are its two future null normal directions. Because the area
density $\rho = \rho_0 v^2$ is not constant along the generators, these future null directions cannot be
rescaled isometrically. The isometry must preserve not only the directions but the vectors $\di_{v_A}$.
In sum, the isometry preserves $S_0 \cap U$ and a complete basis of spacetime vectors at $S_0 \cap U$.
The isometry is therefore trivial, that is $\xi = 0$, throughout $U$.\footnote{
Isometries are rigid in any connected spacetime with a smooth non-degenerate metric: They are
completely determined by their actions on one point of the spacetime and on the tangent space at that
point. See \cite{Wald} p. 442 for a proof.}
This implies in particular that
$\dg \theta^p - \tilde\dg \theta^p = 0$ at $S_A$, so $\dg s_A = \tilde\dg s_A$.

Extending the preceeding argument one can conclude that the condition (\ref{auxbracketdef}),
$\dg A = \omega_\cN[\{A,\cdot\}_\bullet, \dg]$ $\forall \dg \in L_g^0$, 
does not define, nor impose any restriction on, the brackets of the diffeomorphism data,
because brackets involving these data 
do not enter the condition:
Because the observable $A$ is diffeomorphism invariant by definition $\{A,\cdot\}_\bullet$
does not depend on $\{s,\cdot\}_\bullet$ or $\{\bar{v},\cdot\}_\bullet$, and because
$\dg g_{ab}$ vanishes in a neighborhood of $\di\cN$, $\omega_\cN[\{A,\cdot\}_\bullet,\dg]$
does not depend on $\{A,s\}_\bullet$ or $\{A,\bar{v}\}_\bullet$. 
The fact that the variation of the diffeomorphism data are determined by those of the $v$ data
under $\dg \in L_g^0$, and the gauge invariance of $A$, then imply that $\dg s$ and $\dg \bar{v}$
do not enter (\ref{auxbracketdef}) either.  

How was it then possible to obtain the brackets of the diffeomorphism data in \cite{PRL}?
In \cite{PRL} brackets were obtained for all the data, including the diffeomorphism
data, by imposing a strengthened version of (\ref{auxbracketdef}). The bracket was required to
satisfy the conditions
\bearr	\label{auxbracketdefmod}
\dg A & = & \omega_\cN[\{A,\cdot\}_\bullet, \dg]\ \ \forall \dg \in C\\
\{A,\cdot\}_\bullet & \in & C,
\eearr
$C$ being a subset of the admissible variations\footnote{
In \cite{PRL} the term ``admissible variation`` is defined more narrowly than here and
refers only to the variations in $C$.}
containing the null sheet preserving variations in $L_g^0$ as a proper subset.\footnote{
Recall that $\dg$ in (\ref{auxbracketdef}) may be restricted to null sheet preserving variations
in $L^0_g$ without weakening this condition, so the fact that $C$ contains all these variations
implies that (\ref{auxbracketdefmod}) is at least as strong as (\ref{auxbracketdef}).}
\footnote{In \cite{PRL} one natural condition on the bracket is {\em relaxed}, namely the
requierment that the changes in the metric on $\cN$ be real. But the complex variations of 
the metric that are generated via the resulting bracket are special modes that represent 
shock waves that propagate along $\cN$ and do not affect the metric on the interior of
the domain of dependece of $\cN$. A similar relaxation of the reality conditions on the data 
probably has to be made also to obtain a Poisson bracket satisfying (\ref{auxbracketdef}).
This ultimately seems to be a consequence of insisting on defining the Poisson bracket on all 
modes of the initial data, including these schock wave modes which are superflous for describing 
the interior of the domain of dependence.}  
These conditions define an essentially unique bracket on all the data. Of course any bracket
satisfying the stronger condition (\ref{auxbracketdefmod}) also satisfies the weaker condition
(\ref{auxbracketdef}), so the brackets of the $v$ data given in \cite{PRL} are a solution to
(\ref{auxbracketdef}).

\subsection{The $a$ and $b$ charts}	\label{charts}

Two types of special spacetime charts, called ``$a$" charts and ``$b$" charts, will be
used.
The charts $b_L$ and $b_R$ extend the $v\theta$ charts on $\cN_L$ and $\cN_R$ to
charts on an open spacetime neighborhood of the interior, $S_0 -\di S_0$, of $S_0$.
Both are formed from the same coordinates $v^L, v^R, \theta^1,\theta^2$, but they
differ in the ordering of these coordinates: $b_R^\mu = (v^L, v^R, \theta^1,\theta^2)$
and $b_L^\mu = (v^L, v^R, \theta^2,\theta^1)$. That is, the roles of $v^L$ and
$v^R$ are interchanged in the two charts, as are those of $\theta^1$ and $\theta^2$,
so that the charts have the same orientation. The coordinates $v^L$, $v^R$, $\theta^1$,
and $\theta^2$ are obtained from the $v\theta$ charts by setting $v^R = 1$ on $\cN_L$
and $v^L = 1$ on $\cN_R$, and then extending the functions $v^L$, $v^R$, $\theta^1$,
and $\theta^2$ arbitrarily, but smoothly, off $\cN$.  Lowercase indices $\mu,\nu,...$
from the latter part of the Greek alphabet will represent $b$ coordinate indices.

The $a_A$ chart, associated with the branch $\cN_A$, is defined in much the same way
as the $b_A$ chart, but with the truncating 2-surface $S_A$ playing the role of $S_0$.
It consists of the ordered coordinates $a_A^\ag = (u_A, r_A, y_A^1, y_A^2)$. $y^1$
and $y^2$ are constant on the generators of $\cN_A$ and coincide on $S_A$ with the
fixed $y$ chart already introduced to define the diffeomorphism datum $s$.\footnote{
To lighten notation the branch index, $A$, will usually be supressed when there
is little risk of confusion.}
$r$ is an area parameter along the generators like $v$, but normalized to 1 on
$S_A$, so $r = \sqrt{\rho_y/\bar{\rho}} = v/\bar{v}$, where $\rho_y$ is the
area density on cross sections of $\cN_A$ in the $y$ chart, and $\bar{\rho}$ is
the area density on $S_A$ in this chart. $r$, $y^1$, and $y^2$ are extended off
$\cN_A$ by holding them constant on the null geodesics normal to the equal $r$
cross sections of $\cN_A$ and transverse to $\cN_A$.
Finally $u$ is a parameter along these geodesics set to 0 on $\cN_A$ and chosen
such that $\di_u \cdot \di_r = -1$. Greek lowercase indices $\ag,\bg, ...$ from
the beginning of the alphabet will represent $a$ coordinate indices.

On $\cN_A$ the transformation between the $a$ and $b$ charts is quite simple:
\be	\label{btoa_transformation}
r = v/\bar{v}(\theta)\ \ \ 
y^i = s^i(\theta)\ \ \
u = 0.
\ee
In the $a$ chart the spacetime line element at $\cN_A$ takes the form\footnote{
The $e_{ij}$ are the $y$ chart components of the conformal 2-metric, which is
a (2-dimensional) weight $-1$ tensor density.}
\be             \label{a_line_element}
ds^2 = -2du dr + h_{ij}dy^i dy^j = -2du dr + r^2 \bar{\rho} e_{ij}dy^i dy^j.
\ee
The spacetime line element at $S_0$ is also simple in the $b$ chart. It is
\be             \label{b_line_element_S_0}
ds^2 = 2 \chi dv^L dv^R + h_{pq}d\theta^p d\theta^q,
\ee
with $\chi = \di_{v^L} \cdot \di_{v^R}$.


It will be necessary to have control over the orientations of the charts we 
have defined. The sign of the integral of a form over a manifold depends on the
orientation of the manifold.\footnote{
Two overlapping charts are said to be coherently oriented if the transformation between
them has positive Jacobian determinant. The orientation of a manifold is defined by the
choice of a coherently oriented atlas on the manifold. (A manifold may or may not admit
a coherently oriented atlas. If it does it is orientable, and if it does not it is
non-orientable.) See \cite{CDD}.
The orientation of a chart on an oriented manifold is said to be positive, or to match
that of the manifold, if the chart is coherently oriented with the atlas that defines
the orientation of the manifold.}
Given this orientation the integral of the form can be reduced
to an iterated definite integral by choosing a chart $x$ oriented coherently 
with the manifold, and expressing the integrand as a multiple of the coordinate
volume form: $f\,dx^1 \wedge dx^2 \wedge ...\wedge dx^n$, where $f$ is a 
suitable function. The integral is then $\int f dx^1 dx^2 ... dx^n$, with 
the integration over each $x^s$ running from lesser to greater values of $x^s$. 
(See \cite{CDD}.)

An orientation will be chosen, once and for all, for spacetime (or at least a 
neighborhood of $\cN$). Which orientation is chosen does not matter, because 
whichever choice is made, the sign of the 4-volume form $\veg$ will be chosen 
so that its integral over a spacetime region is positive. The value of the 
action (\ref{EH}) is thus independent of the orientation of spacetime chosen.      

The $a$ charts will be positively oriented, that is, their orientations will be 
chosen to match that of spacetime. This can be achieved by choosing a suitable 
orientation for the $y_A$ chart on $S_A$. 

We shall take $\cN$ to be future oriented. It is with this convention that 
equation (\ref{auxbracketdef}) ensures that the Poisson bracket on initial 
data reproduces the Peierls bracket \cite{MR}.
A chart $x^1, x^2, x^3$ on a non-timelike hypersurface is future oriented 
if a spacetime chart, $t, x^1, x^2, x^3$, formed from the $x$ coordinates 
and a time coordinate $t$ which is constant on the hypersurface and 
increasing toward the future, is positively oriented. 

If $r$ increases toward the future on $\cN_A$ then $u$ must also, because
$g_{ur} = -1$ and the metric is assumed to have signature ${}-+++$. $u$ may 
therefore be taken as the time coordinate in the preceeding definition, and
$u, r, y^1, y^2$ as the combined positively oriented spacetime chart. Thus,
in this case, $r, y^1, y^2$ is a future oriented chart on $\cN_A$, and
therefore matches the orientation of this manifold. By a similar argument,
if $r$ decreases toward the future, then $-u, -r, y^1, y^2$ is a positively
oriented chart with $-u$ a time coordinate increasing toward the future, so
$-r, y^1, y^2$ is a future oriented chart on $\cN_A$.
In either case the coordinate along the generators of the future oriented
chart increases from $S_0$ to $S_A$. Thus
\be		\label{integral_3_form}
\int_{\cN_A} f dr \wedge dy^1 \wedge dy^2 = \int_{S_A} d^2 y \int_{r_0}^1 dr f,
\ee 
where the $y$ integrals run from lesser to greater values of these coordinates, 
or, equivalently, $d^2 y$ is interpreted as the positive euclidean coordinate
measure on $S_A$ defined by the $y$ chart. On each generator $r_0 = 1/\bar{v}$ 
is the value of $r$ at $S_0$.

A future orientation can be defined on the two dimensional cross sections
of $\cN_A$ in an entirely analogous manner, with $\cN_A$ now playing the 
role of spacetime in the preceeding definition. The $y$
chart gives precisely this future orientation to $S_A$ if $\cN_A$ is future 
oriented in spacetime. This is also the orientation that $S_A$ has as part of 
the boundary of $\cN_A$.

The $\theta_A$ chart will be oriented coherently with the $y_A$ chart. 
Therefore, if $S_0^{(A)}$ is $S_0$ oriented coherently with $\di\cN_A$,
and thus past oriented with respect to $\cN_A$, then
\be		\label{integral_2_form}
\int_{S_0^{(A)}}  d\theta_A^1 \wedge d\theta_A^2 = -\int_{S_0} d^2\theta.
\ee 
Since $S_0^{(L)}$ and $S_0^{(R)}$ have opposite orientations it follows
at once that the charts $\theta_R$ and $\theta_L$ must be oppositely oriented.
If $(\theta^1,\theta^2)$ is coherently oriented with $y_R$ then
$\theta_R^p = (\theta^1,\theta^2)$ and $\theta_L^p = (\theta^2,\theta^1)$
satisfy our requirements. These correspond to the $b$ charts
$b_R^\mu = (v^L, v^R, \theta^1,\theta^2)$ and
$b_L^\mu = (v^R, v^L, \theta^2, \theta^1)$ defined earlier. The use of
these two $b$ charts, instead of just one, $b_R$ say, makes possible
a completely symmetrical treatment of the two branches.

\subsection{The symplectic potential in terms of the free null data}\label{symp_pot}

According to (\ref{symppotential}) the contribution to the symplectic potential 
of a branch $\cN_A$ of $\cN$ is
\be	\label{symppotentialN_A}
\Theta_A[\dg] =  -\frac{1}{8\pi G}\int_{\cN_A} \dg\Gamma^{[c}_{c b} g^{a]b}
                         \:\varepsilon_{a\cdot\cdot\cdot}, 
\ee
with the whole symplectic potential given by $\Theta_\cN = \Theta_L + \Theta_R$.
Our task is to rewrite $\Theta_{\cN}[\dg]$ in terms of our free initial data 
for admissible variations $\dg$. Taking the curl of this potential then yields the
symplectic 2-form in terms of these data and variations.


In the following only $\Theta_R$ will be computed explicitly. $\Theta_L$ is
entirely analogous, except that $\tau$ is replaced by $-\tau$ because exchanging
$L$ and $R$ in the definition (\ref{twist}) of $\tau$ produces an expresion equal to
$-\tau$.

It will be convenient to decompose the variation $\dg$ into the sum of a
diffeomorphism generator ${\pounds_{\xi}}$ that accounts for the displacement
of the $a_R$ chart under $\dg$, and a variation $\dg^a = \dg - {\pounds}_{\xi}$,
that leaves this $a$ chart fixed. As is explained in detail in appendix \ref{variations},
$\dg^a g$ is the part of the variation of the metric arising from changes of the
metric components in the $a$ chart: In this chart $[\dg^a g]_{\ag\bg}(a) = \dg [g_{\ag\bg}(a)]$.
The remainder, ${\pounds_{\xi}} g$, is of course the part of the variation arising from
the shift of the $a$ chart.
The corresponding decomposition of $\Theta_R$,
\be             \label{decompTheta}
\Theta_R[\dg] = \Theta_R[\dg^a] + \Theta_R[{\pounds}_\xi].
\ee
neatly separates the contribution from the variations of the bulk datum,
the conformal 2-metric $e$, and the variations of the surface data on $S_0$.

$\Theta_R[\dg^a]$ depends only on the variation of $e$. Indeed, in the $a$ 
chart the metric at $\cN_R$ is restricted to the form 
        \be	\label{a_metric}
\begin{array}{cc} g_{\ag\bg} = \left[\begin{array}{cc} \begin{array}{rr} 0 & -1 \\ -1 & 0 
\end{array} & \begin{array}{rr} {} & {} \\ {} & {} \end{array}\\ 
\begin{array}{rr} {} & {} \\ {} & {} \end{array} & h_{ij} \end{array}\right],     &
                  g^{\ag\bg} = \left[\begin{array}{cc} \begin{array}{rr} 0 & -1 \\ -1 & 0 
\end{array} & \begin{array}{rr} {} & {} \\ {} & {} \end{array}\\ 
\begin{array}{rr} {} & {} \\ {} & {} \end{array} & h^{ij} \end{array}\right],     \end{array}
        \ee        
with $h^{ij}$ the inverse of $h_{ij}$. (See (\ref{a_line_element}).) 
Furthermore $h_{ij} = r^2 \bar{\rho} e_{ij}$. Since $\bar{\rho}$, the $y$ chart 
area density on $S_R$, is invariant under admissible variations, $e_{ij}$ is the
only degree of freedom that can vary in $g_{\ag\bg}$.

$\Theta_R[{\pounds_{\xi}}]$, on the other hand, is a surface integral: 
By (\ref{diffTheta1})
\be	\label{diffTheta_R1}
	\Theta_R[{\pounds_{\xi}}] = -\frac{1}{16\pi G} \int_{\di\cN_R} 
			\nabla^a \xi^b\:\veg_{ab\cdot\cdot}. 
\ee
In fact the integral reduces to one over $S_0$, because the integrand vanishes 
elsewhere: Since $\dg^a$ preserves the $a$ chart $\xi$ is determined by
\be
\dg a^\ag = \dg^a a^\ag + {\pounds}_\xi a^\ag = \xi^\ag.
\ee
On $S_R$ both the coordinates $a^\ag$ and their gradients are invariant under 
admissible variations, so $\xi$ and $\nabla\xi$ vanish there. There is thus no 
contribution to (\ref{diffTheta_R1}) from $S_R$. The fact that the contribution
from $\di\cN_R - S_R - S_0$ also vanishes is most easily understood by expressing
the integrand of (\ref{diffTheta_R1}) on this surface in terms of $a$ chart 
components. The pullback of $du$ to $\di\cN_R - S_R - S_0$ vanishes, and those of
$dy^1$ and $dy^2$ are linearly dependent, so the pullback of 
$\nabla^a \xi^b\:\veg_{ab\cdot\cdot}$ is equal to the pullback of 
$2\nabla^{[u} \xi^{i]}\:\veg_{uirj} dr\wedge dy^j$. But 
\be
\nabla^{[u} \xi^{i]} = g^{u\ag}g^{i\bg}\nabla_{[\ag} \xi_{\bg]} = - h^{ik}\di_{[r} \xi_{k]}.
\ee
Since admissible variations are null sheet preserving, $\dg u = 0$ on $\cN_R$. Thus
$\xi_r = - \xi^u = 0$, and it follows that $\di_k \xi_r = 0$. Since admissible variations
preserve the generators on $\di\cN_R - S_R - S_0$, $\dg y^i = 0$ there. 
Thus $\xi_k = h_{ki} \xi^i= 0$, and $\di_r \xi_k = 0$, which establishes the claim.
The diffeomorphism term is therefore
\be	\label{diffTheta_R2}
	\Theta_R[{\pounds_{\xi}}] = -\frac{1}{16\pi G} \int_{S_0^{(R)}} 
			\nabla^a \xi^b\:\veg_{ab\cdot\cdot}, 
\ee
where $S_0^{(R)}$ is $S_0$ oriented coherently with $\di\cN_R$.

The form of the vector field $\xi$ can be restricted quite a bit by gauge 
fixing the variations further. In particular one can ensure that the variations
leave the $b$ chart fixed in a spacetime neighborhood of $S_0 - \di S_0$. First
one adds a diffeomorphism generator $\pounds_w$ to each variation so that 
the generators, and the $v$ parameter on these, are invariant under the total
variation within some neighborhood of $S_0$. This can be achieved with a diffeomorphism
$\pounds_w$ which is pure gauge, that is, without affecting the value of the symplectic
2-form: Recall that admissible variations already preserve the double null sheet
character of the fixed manifold $\cN$, and map the generators in $\di\cN$ to themselves.
Thus $w$ is tangent to $\cN$ and to the generators on $\di\cN$ within a neighborhood of
$S_0$. We will set $w$ to zero in neighborhoods of $S_L$ and $S_R$ and tangent to the
generators on $\di\cN$ wherever it is non-zero. Then the change in the symplectic
potential due to the addition of $\pounds_w$, $\Theta_\cN[{\pounds_{w}}]$, vanishes, on
$S_L$ and $S_R$ because $w$ vanishes in a neighborhood of these surfaces, and on
$\di\cN - S_L - S_R$ by the argument of the preceeding paragraph. Alternatively,
one may note that the addition of $\pounds_w$ to $\dg$ does not affect $\dg^a$ but does
transform $\xi \rightarrow \xi + w$. Since the new variation $\dg + \pounds_w$ is still
admissible, the contribution to $\Theta_R[{\pounds_{\xi}}]$ from $\di\cN_R - S_R - S_0$ remains
zero. The contribution from $S_0$, (\ref{diffTheta_R2}), is affected by the addition of
of $w$ to $\xi$, but the sum, 
\be	\label{diffTheta_sum}
	\Theta_R[{\pounds_{\xi_R}}] + \Theta_L[{\pounds_{\xi_L}}] = \frac{1}{16\pi G} \int_{S_0^{(R)}} \nabla^a [\xi_L - \xi_R]^b\:\veg_{ab\cdot\cdot},
\ee
is not because $w$ cancels out in the difference $\xi_L - \xi_R$. (The minus sign is due
to the fact that the orientation of $S_0^{(L)}$ is opposite to that of $S_0^{(R)}$.)

To ensure that the $b$ chart is fixed under the gauge fixed variations
the $\theta^1$, $\theta^2$ must be set equal to fixed coordinates on $S_0$.
(Recall that the choice of the $\theta$ chart is a gauge degree of freedom
in our formalism, additional to the spacetime diffeomorphism gauge freedom.)
This fixes the $b$ coordinates on $\cN$ in a neighborhood of $S_0$. They may
then be smoothly extended to a fixed chart on a spacetime neighborhood of the
interior of $S_0$.
 
Once this gauge fixing has been carried out $\xi$ simply measures the 
variation of the transformation from the $b$ chart to the $a$ chart
(within the domain in which the $b$ chart is fixed). On $\cN_R$ the 
transformation between these charts is (\ref{btoa_transformation})
\be\label{btoa_transformation2}
r = v/\bar{v}(\theta),\ \ \
y^i = s^i(\theta),\ \ \
u = 0.
\ee
Thus 
\be		\label{xi_a_chart}
\xi = \dg r \di_r + \dg y^i \di_{y^i} = - r \dg \ln \bar{v}\, \di_r + \dg s^i \di_{y^i},
\ee
which is completely determined by the variations of the $S_0$ data
$s^i$ and $\bar{v}$. Recall that $\xi$ is tangent to $\cN_R$, since
admissible variations preserve $u = 0$ on $\cN_R$.

Let us evaluate the bulk term,
\be             \label{Theta_bulk}
\Theta_R[\dg^a] = -\frac{1}{8\pi G}\int_{\cN_R} \dg^a\Gamma^{[\ag}_{\ag\cg} g^{\bg]\cg}
		   \veg_{\bg\cdot\cdot\cdot}
\ee,
in the symplectic potential in terms of the initial data. It is convenient
to use the $a$ chart, since the $a$ coordinates are fixed under $\dg^a$, and
the $a$ components of the variation under $\dg^a$ of a field is simply the
variation under $\dg$ of the $a$ components of the field. Thus for example
\be
[\dg^a\Gamma]^\ag_{\bg\cg}(a) = \dg[\Gamma^\ag_{\bg\cg}(a)],
\ee
where the variation $\dg$ on the right hand side is of the connection coefficients
$\Gamma^\ag_{\bg\cg}$ evaluated at a fixed $a$ coordinate point $a$.

The form (\ref{a_metric}) of the metric in the $a$ chart implies that 
$\sqrt{-g} = \sqrt{det[h_{ij}]} \equiv \rho_y$ and therefore, since the
$a$ chart is positively oriented, that 
$\veg = \rho_y\, du\wedge dr \wedge dy^1 \wedge dy^2$.
Pulling the last three indices of this 4-volume form back to $\cN_R$ one obtains
\be
\veg_{\bg\cdot\cdot\cdot} = \rho_y\, \dg_\bg^u\, dr \wedge dy^1 \wedge dy^2.
\ee

The integrand in (\ref{Theta_bulk}) thus reduces to $dr \wedge dy^1 \wedge dy^2$ 
times the function
\bearr
\dg^a\Gamma^{[\ag}_{\ag\cg} g^{u]\cg} \rho_y
& = & \frac{1}{2}\{ - \dg^a\Gamma^{\ag}_{\ag r} + g^{\ag\cg} \dg^a\Gamma_{r\ag\cg}\}\rho_y\\
& = & - \frac{1}{2}\{ \dg^a\di_r \ln \sqrt{-g} + 2 \dg^a\Gamma_{rur}
- h^{ij} \dg^a\Gamma_{rij}\}\rho_y\\
& = & - \frac{1}{2}\{ \dg^a\di_r \ln \rho_y + \dg^a\di_u g_{rr} 
+ \frac{1}{2}h^{ij} \dg^a\di_r h_{ij}\}\rho_y.    \label{integrand_Theta_bulk}
\eearr

Now recall that $\rho_y(y,r) = r^2\bar{\rho}(y)$ and that $\dg^a \bar{\rho} = 0$ for admissible variations.
It follows that
\be     \label{dirrho0}
\dg^a\di_r\ln\rho_y = \dg^a 2/r = 0. 
\ee
and
\be     \label{dirhh}
\frac{1}{2}\,h^{ij}\dg^a[\di_r h_{ij}] 
= \dg^a\di_r \ln\rho_y - \frac{1}{2}\di_r[\rho_y\, e_{ij}] \dg^a\frac{e^{ij}}{\rho_y} 
= - \frac{1}{2}[\di_r e_{ij}] \dg^a e^{ij}.
\ee
(Here $e^{ij}$ is the inverse of $e_{ij}$,\footnote{
The $e^{ij}$ are the $y$ coordinate components of the inverse conformal 2-metric defined
earlier, which is a weight $1$, 2-dimensional tensor density.}
and the fact that, for any variation
$\Dg$, $e_{ij}\Dg e^{ij} = -\Dg \ln \det[e_{ij}] = - \Dg 0 = 0$ has been used.)

The remaining, middle, term in (\ref{integrand_Theta_bulk}) is proportional to 
$\dg^a \Gamma^r_{rr}$, since on $\cN_R$
\be     \label{dga_diu_grr}
        \di_u g_{rr} = 2 \Gamma^r_{rr}.
\ee

Because $\cN_R$ is oriented toward the future the form integral (\ref{Theta_bulk})
reduces to an iterated definite integral according to the formula 
(\ref{integral_3_form}). Substituting in our results on the integrand one obtains 
\be
\Theta_R[\dg^a] = -\frac{1}{16\pi G} \int_{S_R} d^2 y\,\bar{\rho} \int_{r_0}^1 
\frac{1}{2} r^2 \di_r e_{ij}\dg^a e^{ij} - 2 r^2 \dg^a\Gamma^r_{rr} dr, 
\ee
where $d^2y$ is the positive euclidean coordinate measure associated with the 
$y$ chart. The first term in the integrand is expressed in terms of the bulk 
datum $e$. The second term combines with a term in the surface contribution
to the symplectic potential to form a total variation, which may be dropped from
the potential without affecting the symplectic 2-form.

Let us now turn to the surface term, $\Theta_R[{\pounds_{\xi}}]$, in
the symplectic potential. Here it is convenient to work with the $b$ chart, 
since the $S_0$ data is defined in terms of this chart. 

Our first task will be to calculate the 4-volume form in the $b$ chart.
Because $n_L = \di_{v^L}$ and $n_R = \di_{v^R}$ are null and normal to $S_0$ 
the spacetime line element at $S_0$ takes the form
(\ref{b_line_element_S_0})
\be
ds^2 = 2 \chi dv^L dv^R + h_{pq}d\theta^p d\theta^q,
\ee
with $\chi = n_L\cdot n_R$. But the sign of $\chi$ depends on the direction in
which $v^L$ and $v^R$ increase. Recall the definition $\sg_A = 1$ if $v^A$
increases to the future, and $-1$ if it decreases. Then  $\sg_A n_A$ is
future directed and $\sg_L\sg_R \chi$ is negative. Indeed $\chi = - \sg_L\sg_R\,e^{-\lam}$.

The chart $(\sg_L v^L, \sg_R v^R, \theta^1, \theta^2)$ has the same, 
positive, orientation as the $a_R$ chart, so the 4-volume form is
\bearr
\veg & = &  |\chi|\, \rho_0\, \sg_R\sg_L\, dv^L \wedge dv^R \wedge d\theta^1 \wedge d\theta^2\\
	 & = &  -\chi\, \rho_0\, dv^L \wedge dv^R \wedge d\theta^1 \wedge d\theta^2\\
     & = &  -\frac{\rho_0}{\chi} n_R \wedge n_L \wedge d\theta^1 \wedge d\theta^2.
\eearr
In the last line the $n_A$ denote the 1-forms $n_{L\,a} = \chi\di_a v^R$ and 
$n_{R\,a} = \chi\di_a v^L$, obtained by lowering the indices of the tangent 
vectors $n_A^a$ with the metric.

In (\ref{diffTheta_R2}) the first two indices of $\veg_{abcd}$ are contracted and 
the last two indices are pulled back to $S_0$. When thus pulled back $\veg_{ab\cdot\cdot}$
becomes
\be
\veg_{ab\cdot\cdot} = -2 \frac{\rho_0}{\chi}\, 
	n_{R\,[a} n_{L\,b]}\, d\theta^1 \wedge d\theta^2.
\ee 
Substituting this expression into (\ref{diffTheta_R2}) yields
\be     \label{ThetaR0}
\Theta_R[{\pounds_\xi}] = \frac{1}{16\pi G}\int_{S_0} \frac{1}{\chi} 
\{n_R \cdot\nabla_{n_L} \xi - n_L \cdot \nabla_{n_R} \xi\}\, \rho_0\, d^2\theta, 
\ee
with $d^2\theta$ the positive euclidean measure defined by the $\theta$ chart 
on $S_0$. The formula (\ref{integral_2_form}), which takes into account the 
orientation of $S_0$ employed in (\ref{diffTheta_R2}), has been used to turn 
(\ref{diffTheta_R2}) into a definite integral.

The derivative along $n_L$ may be eliminated in favour of a variation of $\chi$:
\bearr
\dg\chi = n_L^a n_R^b \dg g_{ab} & = & n_L^a n_R^b[\dg^a g_{ab} + {\pounds_\xi}\, g_{ab}] \\
& = &  n_L^\ag n_R^\bg\, \dg^a g_{\ag\bg} + 2 n_L^a n_R^b \nabla_{(a} \xi_{b)}.
\eearr
But $n_R^\bg= \di_{v^R} a^\bg = 1/\bar{v}\, \dg_r^\bg$, and (by (\ref{a_metric})) 
$g_{\ag r} = -\dg^u_\ag$, which is of course invariant, so
$n_L^\ag n_R^\bg\, \dg^a g_{\ag\bg} = 0$ and
\be
\dg\chi = n_R\cdot\nabla_{n_L}\xi + n_L\cdot\nabla_{n_R}\xi. 
\ee
The integrand of (\ref{ThetaR0}) is thus equal to
\be             \label{integrandThetaR0}
\frac{1}{\chi}[ \dg\chi - 2 n_L\cdot\nabla_{n_R}\xi ]\rho_0 
= -\rho_0 \dg\lam - 2\frac{\rho_0}{\chi}n_L\cdot\nabla_{n_R}\xi,
\ee
since $\lam = -\ln |\chi|$.

Notice that the $\theta$ components of $\xi$, 
$\xi^p = \xi \cont d\theta^p = \dg s^i\, \di_{y^i}\theta^p$, are independent of 
$v$. This means that $\xi_\perp \equiv \xi^p \di_p$ is Lie dragged along 
$n_R = \di_v$: 
\be
0 = {\pounds}_{n_R} \xi_\perp = \nabla_{n_R} \xi_\perp - \nabla_{\xi_\perp} n_R, 
\ee
for in the $b$ chart the Lie derivative along $n_R$ reduces to simply the $v$ 
partial derivative of the $b$ components of $\xi_\perp$.

The second term in (\ref{integrandThetaR0}) may therefore be expanded according to
\be     \label{integrandTheta1}
2 n_L \cdot \nabla_{n_R} \xi = 2 n_L \cdot \nabla_{n_R}[\xi_\perp + \xi^v n_R] 
=  2 n_L\cdot \nabla_{\xi_\perp} n_R + 2 \xi^v n_L \cdot \nabla_{n_R} n_R + 2\chi d_{n_R} \xi^v   
\ee
The first term in this expansion is linear in the twist $\tau$.
By the definition (\ref{twist}) of $\tau$ 
\be     \label{first_term}
2 n_L\cdot \nabla_{\xi_\perp} n_R = d_{\xi_\perp} \chi + \chi \xi_\perp \cont \tau
= - \chi \xi_\perp \cont [d\lam - \tau].
\ee
The second and third terms in (\ref{integrandTheta1}) are proportional to the values on $S_0$ of 
$\xi^v$ and $d_{n_R}\xi^v = \di_v \xi^v$ respectively. In a neighborhood of $S_0$ $\dg v = 0$, 
so $\dg^a v = - {\pounds}_\xi v = - \xi^v$ there. It follows that   
\be
  \xi^v = -\dg^a v = -\dg^a [r/r_0] = v\, \dg^a \ln r_0,
\ee
and therefore on $S_0$
\be	\label{xi_v_S_0}
\xi^v = \di_v\xi^v = \dg^a \ln r_0.
\ee
The second term in (\ref{integrandTheta1}) also contains a factor
$n_L \cdot \nabla_{n_R} n_R$. Since $n_R$ is tangent to a geodesic and
$n_R = \di_v = r_0\di_r$
\be
n_L\cdot\nabla_{n_R} n_R = n_L \cdot r_0^2 \nabla_{\di_r} \di_r 
= n_L\cdot r_0 \Gamma^r_{rr} n_R = \chi r_0 \Gamma^r_{rr}.
\ee
The second and third terms are the ones that will combine with terms from the 
bulk contribution to the symplectic potential to form a total variation. 

Substituting our results into (\ref{ThetaR0}) and adding the bulk contribution
$\Theta_R[\dg^a]$ we obtain the complete symplectic potential of $\cN_R$: 
\bearr
\Theta_R[\dg] & = & -\frac{1}{16\pi G}\bigg\{\int_{S_0} [\dg \lam - \xi_{\perp} \cont (d \lam - \tau)
+ 2 \dg^a \ln r_0 (1 + r_0\Gamma^r_{rr}) ]\rho_0\,d^2 \theta \nonumber \\
&& + \int_{S_R} d^2 y\,\bar{\rho} \int_{r_0}^1 
[\frac{1}{2} r^2 \di_r e_{ij}\dg^a e^{ij} - 2 r^2\dg^a\Gamma^r_{rr}] dr
\bigg\} \label{symp_potential3}
\eearr

The dependence on $\Gamma^r_{rr}$ can be eliminated. As has already been pointed
out, adding the variation of a functional of the data to the symplectic 
potential does not affect its curl, the symplectic 2-form. Thus we are free
to subtract from $\Theta_R[\dg]$ the variation
\bearr
\lefteqn{\dg \left[ \frac{1}{8\pi G} \int_{S_R} d^2 y\,\bar{\rho} \int_{r_0}^1 
[r + r^2\Gamma^r_{rr}] dr \right]} \nonumber\\
& = & \frac{1}{8\pi G} \left\{\int_{S_R} d^2 y\,\bar{\rho} \int_{r_0}^1 r^2 \dg^a\Gamma^r_{rr} dr 
- \int_{S_0} d^2 \theta\,\rho_0\,\dg^a\ln r_0\, (1 + r_0\Gamma^r_{rr})\right\}
\label{var_expansion}.
\eearr
Recall that the variation $\dg [F]_a$ of the $a$ chart
components of a field $F$, at a fixed $a$ coordinate point, is given by the $a$
chart components of $\dg^a F$. This is the reason for the appearance of
$\dg^a$ in (\ref{var_expansion}). (See (\ref{delta_int_a2}) of appendix \ref{variations}.)
Use has also been made of the fact that $r_0^2\bar{\rho}$ is the $y$ chart area
density on $S_0$, so $r_0^2\bar{\rho}\, d^2y = \rho_0\, d^2\theta$.

We will therefore take as the symplectic potential of $\cN_R$
\be	\label{symp_potential4}
\Theta'_R[\dg] = -\frac{1}{16\pi G}\left\{ 
\int_{S_0} [\dg \lam - \xi_{\perp}\cont(d\lam - \tau)]\rho_0\, d^2\theta +
\frac{1}{2}\int_{S_R} d^2 y\, \bar{\rho} 
	\int_{r_0}^1  r^2 \di_r e_{ij} \dg^a e^{ij} dr 
\right\}.
\ee
Proceeding in exactly the same way an analogous expression is obtained for
$\Theta'_L$, the symplectic potential of $\cN_L$, with the one difference that 
$\tau$ is replaced with $-\tau$ since interchanging $L$ and $R$ maps $\tau$ to
$-\tau$.

Equation (\ref{symp_potential4}) and its $L$ branch analog provide an expression 
for the symplectic potential entirely in terms of our free null initial data. 
It depends on the $v$ data $\rho_0$, $\lam$, and $\tau$ on $S_0$, and on
$\xi_{\perp\,A} = \dg s_A^i\, \di_{y_A^i}$ there. It further depends on the $a_A$ chart 
conformal 2-metric $e_{ij}$ on $\cN_A$, on the (invariant) $y_A$ chart area density
$\bar{\rho}_A = |det \frac{\di s_A^i}{\di\theta^p}|^{-1}\rho_0 \bar{v}_A^2$ on $S_A$, 
and on $r_{A\,0} = 1/\bar{v}_A$. Note that the transformation from the $b$ chart conformal
metric $e_{pq}(v,\theta)$, which is one of our data, to $e_{ij}(r,y)$ is determined by the
transformation (\ref{btoa_transformation2}) from the $b$ chart to the $a_A$ chart on 
$\cN_A$, which in turn is determined by the diffeomorphism data $s_A$ and $\bar{v}_A$.

We have achieved our goal of expressing the symplectic potential in terms of the
null initial data of subsection \ref{data}. It turns out however that a symplectic
potential that is in some ways more useful is obtained by replacing the datum $\tau$
by two new data
\be
\tilde{\tau}_{R\,i} \equiv \rho_0  (d\lam - \tau) \cont \di_{y_R^i}\ \ \mbox{and}\ \ 
\tilde{\tau}_{L\,j} \equiv \rho_0  (d\lam + \tau) \cont \di_{y_L^j}.
\ee
These are the coefficients of $\dg s_R^i$ and $\dg s_L^j$ respectively in
the surface term of the symplectic potential. In terms of these new data
\be	\label{symp_potential5}
\Theta'_R[\dg] = -\frac{1}{16\pi G}\left\{ 
\int_{S_0} \rho_0\,\dg \lam - \tilde{\tau}_{R\,i}\dg s_R^i\,d^2\theta 
+ \frac{1}{2}\int_{S_R} d^2 y\, \bar{\rho} 
	\int_{r_0}^1  r^2 \di_r e_{ij} \dg^a e^{ij} dr \right\},
\ee
and $\Theta'_L$ is given by a completely analogous expression. (In particular
$\tilde{\tau}_L$ enters $\Theta'_L$ in precisely the same way as $\tilde{\tau}_R$ enters
$\Theta'_R$, since the difference in the sign with which $\tau$ enters $\Theta'_L$ and
$\Theta'_R$ has been absorbed into the definitions of $\tilde{\tau}_R$ and $\tilde{\tau}_L$.)
In principle $\tilde{\tau}_R$ and $\tilde{\tau}_L$ are related by the equation
\be	\label{tau_constraint}
\tilde{\tau}_{R\,i} ds_R^i + \tilde{\tau}_{L\,j} ds_L^j  = 2\rho_0 d\lam.
\ee
However, we shall extend the phase space by taking $\tilde{\tau}_R$ and 
$\tilde{\tau}_L$ to be independent, and then treat (\ref{tau_constraint}) 
as a constraint which defines our original phase space. This constraint generates the
gauge transformations of the $\theta$ chart \cite{MR}. Thus, in the extended phase
space, without the constraint, these transformations are not gauge, and the symplectic
2-form is in fact non-degenerate.

The introduction of constrained variables seems a step backward with respect 
to our aim of a canonical description in terms of free data, but the constraint 
introduced brings no real complications. Indeed, (\ref{tau_constraint}) may
be solved easily for $\tilde{\tau}_R$. If the $\theta$ chart is then fixed via
the gauge condition $s_R = \mbox{id}$ (i.e. $\theta = y_R$), then the physical 
phase space is parametrized by the remaining data, and the Dirac brackets of 
these remaining data are equal to their brackets in the extended phase space 
\cite{MR}.

A non degenerate symplectic form can only be achieved by either extending the
phase space as we do, or by gauge fixing the $\theta$ chart. The use of an
unfixed, arbitrary, $\theta$ chart, has made it possible to treat the two 
branches of $\cN$ autonomously and symetrically. This is also possible for 
some gauge fixed $\theta$ charts. For instance one could take the $\theta^p$ 
to be isothermal coordinates of the metric on $S_0$. This gauge choice has the
drawback that isothermal coordinates depend non-locally on the metric. As a
result the Dirac bracket, unlike the extended phase space Poisson bracket, 
does not always vanish between data on distinct generators. Other gauge 
fixings which avoid this complication can be defined, but all the same, 
leaving $\theta$ unfixed and working with the extended phase space seems the 
simplest choice. 

\subsection{The symplectic 2-form in terms of the free null data}\label{symp2}

The contribution of the hypersurface $\cN_R$ to the symplectic form is
\be 
\omega_R[\dg_1,\dg_2] = \dg_1 \Theta'_R[\dg_2] - \dg_2 \Theta'_R[\dg_1] 
- \Theta'_R[[\dg_1,\dg_2]].
\ee
Here this expression will be evaluated in terms of the free null initial data for
admissible variations $\dg_1$ abd $\dg_2$. Since $[\dg_1,\dg_2]$ is also admissible
the symplectic potential is needed only on admissible variations. Equation
(\ref{symp_potential5}) for $\Theta'_R$ therefore provides a sufficient basis for
the calculation.

The first term of $\Theta'_R$ in (\ref{symp_potential5}) is a surface term, an 
integral over $S_0$, while the second term is a bulk term, an integral over 
$\cN_R$. As a result $\omega_R$ also consists of bulk and surface terms. The bulk 
term in $\omega_R$ is obtained by varying the bulk term in $\Theta'_R$ with $r_0$ 
held fixed. It is 
\be     
\frac{1}{32\pi G} \int_{S_R} d^2 y\, \bar{\rho} \int_{r_0}^1 r^2 \dg^a_1 e^{ij}
\di_r \dg^a_2 e_{ij}\: dr - (1 \leftrightarrow 2).     \label{omega_bulk_term}
\ee
Since $r_0$ is held fixed the domain of integration does not vary in the $a$ chart.
The variation of the integral is therefore just the integral of the variation of
the integrand in this chart, that is, of $\dg^a$ of the integrand. 
(See (\ref{delta_int_a2}) of appendix \ref{variations}.)

In terms of the $v\theta$ chart the bulk term in $\omega_R$ may be written as
\be     
\frac{1}{32\pi G} \int_{S_0} d^2 \theta\,\rho_0 \int_1^{\bar{v}} v^2 \dg^a_1 e^{pq}
\di_v \dg^a_2 e_{pq}\: dv - (1 \leftrightarrow 2).     \label{omega_bulk_term2}
\ee 
(Note that the transformation between $y$ and $\theta$ components is independent
of $r$, so it may be freely moved through the derivative, $\di_r$, in 
(\ref{omega_bulk_term}).)
 
The surface contribution to $\omega_R$ comes both from the surface term in 
$\Theta'_R$ and from the variation of $r_0$ in the bulk term of $\Theta'_R$. 
The surface term in $\Theta'_R$ yields
\be
\frac{1}{16\pi G}\int_{S_0}
\dg_1 \lam\, \dg_2\rho_0 + \dg_1\tilde{\tau}_{R\,i}\,\dg_2 s_R^i\,d^2\theta
- (1 \leftrightarrow 2). \label{variation_S0_term}
\ee
The variation of $r_0$ in the bulk term in $\Theta'_R$ produces
\bearr
\lefteqn{\frac{1}{32\pi G} \int_{S_R} \dg_1 [r_0]_y [r^2 \di_r e_{ij}
\dg^a_2 e^{ij}]_{r = r_0} \bar{\rho}\, d^2 y - (1 \leftrightarrow 2) }  \nonumber \\ 
& = & \frac{1}{32\pi G}\int_{S_0} \dg_1^a \ln r_0\, \di_v e_{pq}
\dg^a_2 e^{pq} \rho_0\, d^2\theta - (1 \leftrightarrow 2),   \label{r0_variation}
\eearr
where $\dg [r_0]_y = \dg^a r_0$ is the variation of the scalar $r_0(y)$ at constant $y$.
(See (\ref{delta_int_a2}).)

In (\ref{r0_variation}) $\dg^a$ may be replaced by the variation
$\dg^y \equiv \dg - {\pounds}_{\xi_\perp} = \dg^a + {\pounds}_{\xi^v\di_v}$, associated
with the hybrid chart $(v^L, v^R, y^1, y^2)$. Clearly $\dg^a \ln r_0 = \dg^y \ln r_0$ since
$r_0$ only depends on $y$.   
Furthermore, as will soon be demonstrated
\be	\label{Lie_along_generator}
{\pounds}_{\xi^v\di_v}e^{pq} = \xi^v\di_v e^{pq}.
\ee
Substituting these two relations into the integrand of (\ref{r0_variation}), and
taking into account that by (\ref{xi_v_S_0}) $\xi^v = \dg^a \ln r_0 = \dg^y \ln r_0$ on $S_0$,
one obtains
\bearr
\lefteqn{\dg_1^a \ln r_0 \:\dg_2^a e^{pq} - (1 \leftrightarrow 2) }\\
&&  = \{\dg_1^y \ln r_0 \:\dg_2^y e^{pq} - \dg_1^y \ln r_0 \: \dg_2^y \ln r_0 \di_v e^{pq}\}
- (1 \leftrightarrow 2)\\
&&  = \dg_1^y \ln r_0 \:\dg_2^y e^{pq} - (1 \leftrightarrow 2).
\eearr

Equation (\ref{Lie_along_generator}) can be demonstrated as follows:
Any variation $\Delta h_{tu}$ of the 2-metric on $\cN_R$ gives rise to a variation
\be \label{Delta_e_expansion}
\Delta e^{pq} = \Delta [\sqrt{\det h}h^{pq}]
= -h^{pr}h^{qs}\sqrt{\det h}
	[\Delta h_{rs} - \frac{1}{2}h_{rs}h^{tu}\Delta h_{tu}]
\ee
of the inverse conformal 2-metric. But because the $v$ components of the induced
metric on $\cN_R$ vanish
\bearr	
{\pounds}_{\xi^v\di_v}h_{rs} & = & {\pounds}_{\xi^v\di_v}g_{rs}\\
& = & \xi^v\di_v g_{rs} + \di_r \xi^v g_{vs} + \di_s \xi^v g_{rv} = \xi^v\di_v h_{rs}.
\eearr
The result (\ref{Lie_along_generator}) follows directly from this relation and
(\ref{Delta_e_expansion}). (See (\ref{Lie_def}) for a general definition of the Lie derivative.)

Using the invariance of $\bar{\rho}$ the variation $\dg^y \ln r_0$
may be expressed in terms of $\dg^y \rho_0$. 
$r_0^2 = \rho_{y\,0}/\bar{\rho}$, where $\rho_{y\,0}$ is the area 
density on $S_0$ in the $y$ chart, so, since $\dg^y \bar{\rho} 
= \dg^a \bar{\rho} = 0$,  
\be 
\dg^y \ln r_0 = \frac{1}{2}\frac{\dg^y \rho_{y\,0}}{\rho_{y\,0}}.
\ee
But $\dg^y \rho_{y\,0}$ is just $\dg^y \rho_0$ transformed, 
as a density, from the $\theta$ chart to the $y$ chart, so 
$\dg^y \rho_{y\,0}/\rho_{y\,0} = \dg^y \rho_0/\rho_0$.
(See the discussion of the transformation under change of coordinates
of comoving variations, such as  $\dg^y \rho_0$, in appendix \ref{variations}.)
Thus
\be
\dg^y \ln r_0 = \frac{1}{2}\frac{\dg^y \rho_0}{\rho_0}.
\ee
The contribution (\ref{r0_variation}) to $\omega_R$ can therefore be written as
\be
\frac{1}{64\pi G}\int_{S_0} \dg^y_1 \rho_0 \di_v e_{pq}
\dg^y_2 e^{pq}\, d^2\theta - (1 \leftrightarrow 2).        \label{r0_variation2}
\ee

Summing (\ref{omega_bulk_term2}), (\ref{r0_variation2}), and 
(\ref{variation_S0_term}) one obtains
\bearr
\setlength\arraycolsep{0pt}
\omega_R[\dg_1,\dg_2] = \frac{1}{16\pi G}\int_{S_0} d^2\theta && \!\!\!\!\!\!\!\!\!\!
\left\{ \dg_1\lam \dg_2\rho_0
 + \dg_1\tilde{\tau}_{R\,i}\dg_2 s_R^i
 + \frac{1}{4}\dg_1^y \rho_0 \di_v e_{pq} \dg^y_2 e^{pq}\right. \nonumber \\
 && \!\!\!\!\!\!\!\left. + \frac{1}{2} \rho_0 \int_1^{\bar{v}} v^2
\dg^a_1 e^{pq}\di_v \dg^a_2 e_{pq} dv\:\right\} - (1 \leftrightarrow 2),
\label{omega_free_data} 
\eearr
The sum of (\ref{omega_free_data}) and its $L$ branch analog is the 
desired expression for the symplectic 2-form $\omega_\cN = \omega_R + \omega_L$ 
in terms of the free initial data, valid for admissible variations. It coincides with
the expression given in \cite{PRL} (although there $\dg^a e$ was called $\dg^\circ e$).

The variations appearing in (\ref{omega_free_data}) are not simply the variations of the
components of the intial data fields. For instance, $\dg^a e_{pq}(\theta)$ is not the
variation of $e_{pq}(\theta)$ but rather this variation minus $\pounds_\xi e_{pq}(\theta)$.
Expressed directly in terms of the variations of the components of the initial data fields,
each in a chart ``natural'' to it, the symplectic form is
\bearr
\omega_R[\dg_1,\dg_2] = \frac{1}{16\pi G}\bigg\{ &\!\! \frac{1}{2} &
\!\!\!\int_{S_R} d^2 y\, \bar{\rho} \int_{r_0}^1 r^2 
\dg_1 e^{ij}\,\di_r \dg_2 e_{ij}\, dr \ \ \ \ \ \ \ \ \ \nonumber\\
\!\!&\!\! + \frac{1}{4} & \!\int_{S_0} \dg_1\rho_{y\,0}
\:\di_v e_{ij}\, \dg_2 [e^{ij}(y)]\: d^2 y \nonumber\\
\!&\!+ &\!\! \int_{S_0}[\dg_1\lam\dg_2\rho_0  + \dg_1\tilde{\tau}_{R\,i}\,\dg_2 s_R^i]\:
d^2\theta - (1 \leftrightarrow 2)\bigg\}.
\label{omega_R}
\eearr  
This is the expression given in \cite{MR}. In the first, bulk, term the components of $e$
are referred to the $a$ chart; In the second term, a surface term, $e$ and $\rho_0$ are
referred to the $y$ chart on $S_0$; And in the last term $\lam$, $\rho_0$,
$\tilde{\tau}_{R\,i}$, and $s_R^i$ are referred to the $\theta$ chart. Of course the
components $s^i_R$ of $s_R$ are also determined by the $y$ chart. What is meant in this case
is that the variation of these components is evaluated at constant $\theta$.

\section*{Acknowledgments}

I would like to thank Ingmar Bengtsson, Rodolfo Gambini, Carlo Rovelli, Alejandro Perez, 
Laurent Freidel, Lee Smolin, Robert Oeckl and Jose Zapata, for fruitful discussions. I 
would also like to thank the Centre de Physique Th\'{e}orique in Luminy, the Albert Einstein 
Institute in Potsdam, the Perimeter Institute in Waterloo, and the Centro de Ciencias 
Matem\'{a}ticas de la UNAM in Morelia, for their hospitality during the realization of this work.

\appendix

\section{Variations of fields and integrals}\label{variations}


In the main text extensive use is made of charts that are adapted to the metric, in particular
the $a$ and $b$ charts. We call such charts {\em moving charts} because they can change under variations
of the fields. In the present appendix some basic facts about the variations of fields and their components
in such charts are derived.


In our formalism fields will be defined by their components in charts. The components of a field
$F$ in a chart $x$, denoted $[F]_x$, is a collection of numbers, or more precisely, $\mathbb{C}$
valued functions of the coordinates $x^\mu$ of the chart $x$. Of the chart dependence of the components
we will require only that within the domain of $x$ the components of $F$ in any other chart $y$ are
determined entirely by its $x$ components and the transition function from the $x$ to the $y$ chart.
The advantage of this coordinate dependent representation of fields is that it allows us to treat
all the fields that we encounter, tensors, densities, connection coefficients and others, in a
uniform manner.


Recall that by virtue of its definition a manifold comes equiped with an atlas of charts
(see for example \cite{Wald}). We will call these the {\em fixed} charts. A fixed chart
assigns definite values to the coordinates at each manifold point in its domain.
Moving charts are families of fixed charts depending on the values fields or parameters.
The coordinate values they assign to points can depend on these fields or parameters. For
instance, the $a_A$ chart defined in subsection \ref{charts}, depends on the metric and
moves when the metric is varied.


The variation of a function $f$ of a parameter $\lam$ when $\lam$ varies is simply
another word for the derivative of $f$: $\dg f \equiv df/d\lam$. We will be interested
chiefly in the variations induced by variations of the metric field. Thus we have
a family of metric fields parameterized by $\lam$ and we wish to find the derivative
in $\lambda$ of quantities calculated from the metric. 


The variation of a function of $\lam$ only is thus unambigously defined. But a field
depends also on position on the manifold, and its components depend on the chart used.
Thus to define the variation of a field $F$ one must define what it means to hold the
position and the chart constant while $\lam$ is varied. Here we define
$\dg F$ in the usual way, as the $\lam$ derivative of $F$ in a fixed chart.
That is, we set the components $[\dg F]_x$ of $\dg F$ in a fixed chart $x$ equal to the variation
of the $x$ components of $F$ at fixed values of the coordinates $x$:
\be	\label{delta_fixed}
[\dg F]_x = \dg[F]_x.
\ee

Variations may also be defined in an entirely analogous manner using moving charts: Let
$C$ be an atlas of comoving charts, that is, an atlas of charts which are $\lam$ dependent
functions of the fixed charts, but have $\lam$ independent transition functions among
themselves. Then the components of the {\em comoving variation} $\dg^C F$ in any chart
$c \in C$ are
\be	\label{delta_moving}
[\dg^C F]_c = \dg[F]_c,
\ee  
where $\dg[F]_c$ denotes the variation of the $c$ components at fixed values of the
$c$ coordinates. For any given value of $\lam$ the moving chart $c$ coincides with a fixed
chart $c_\lam$, and the components of a field with respect to $c$ are identified with the
$c_\lam$ components of the field.\footnote{
But $\dg[F]_c$ is the derivative $d/d\lam$ of $[F]_c = [F]_{c_\lam}$ holding constant the values of the
$c$ coordinates, not those of the $c_\lam$ coordinates at fixed $\lam$.}
Note that within its domain a moving chart $c$ defines its comoving atlas uniquely, and
thus also the corresponding comoving variation, which may as well be written $\dg^c$.
Thus for instance the variation $\dg^{a_A}$ used in the main text is defined by the $a_A$ chart.
\newline

\noindent {\bf Proposition:}
\be	\label{deltaC_Lie}
\dg^C = \dg + \pounds_v.
\ee
Here $\pounds_v$ is the Lie derivative along the ``velocity'' $v$ of the moving charts in $C$
with respect to the fixed charts. $v^\mu$ is the $\lam$ rate of change of the fixed chart coordinate
$x^\mu$ corresponding to constant values of the coordinates of the moving charts in $C$. Equivalently,
let $\Phi_{C\,\lam}$ be the diffeomorphism such that $c_\lam^\ag(\Phi_{C\,\lam}(p)) = c_0^\ag(p)$
for each chart $c \in C$ and each point $p$ in the domain of $c_0$. Then $v$ is the velocity of the
flow $\Phi_{C\,\lam}$, i.e. the tangent of the curve $\lam \mapsto \Phi_{C\,\lam}(p)$ at $\lam = 0$.

This proposition simply expresses the fact that the variation of $[F]_c$, the moving chart
components of a field $F$, can be resolved into the sum of a variation, $\dg$, holding the chart
fixed, and a variation, $\pounds_v$,  holding the fixed chart components of $F$ fixed. Without
loss of generality we may suppose that the variation is being evaluated at $\lam = 0$. Then
\be	\label{delta_moving2}
[\dg^C F]_{c_0} \equiv \dg[F]_c = d/d\lam([F]_{c_0}) + d/d\lam([F_0]_c),
\ee
where $F_0$ is the field $F$ at $\lam = 0$ in the sense that $[F_0]_x = [F]_x$ at $\lam = 0$ in
any fixed chart $x$. The first term is $[\dg F]_{c_0}$. The second term turns out to be the Lie
derivative of $F$.

The Lie derivative is defined in terms of the action of diffeomorphisms on the field (see \cite{Wald}).
The action $\Phi^*$ of a diffeomorphism $\Phi$ on a field $F$ satisfies the requierment that for any chart $x$
\be
[\Phi^*(F)]_{\Phi^*(x)} = [F]_x,
\ee
where $\Phi^*(x) \circ \Phi = x$. That is, one requiers that if one acts on both the chart and the field
with the same diffeomorphism then the components of the new field in the new chart are the same as
those of the old field in the old chart. Thus $[F]_{c_\lam} = [\Phi_{C\,\lam}^{*\,-1}(F)]_{c_0}$ and
thus at $\lam = 0$
\be	\label{Lie_def}
[\pounds_v F]_{c_0} \equiv - d/d\lam [\Phi_{C\,\lam}^*(F_0)]_{c_0} = d/d\lam [\Phi_{C\,\lam}^{*\,-1}(F_0)]_{c_0}
= d/d\lam [F_0]_{c_\lam},
\ee
which completes the proof of the proposition.

The variations $\dg^C$ may be interpreted in a different way, as variations with respect to the
fixed charts that leave the moveable atlas $C$ fixed. Indeed, since the variation $\dg$ in
(\ref{deltaC_Lie}) determines the vector field $v$, $\dg^C$ may be regarded as a projection of
$\dg$ to variations that fix $C$.\footnote{It is important to remember that $v$ depends on $\dg$.
Thus for instance, if $\dg$ already fixes $C$, then $v = 0$.} 

 
How do the components of $\dg F$ and $\dg^C F$ transform from one chart to another?
Suppose $x$ and $y$ are two fixed charts. Recall that within the intersection of the domains
of these charts the $y$ components of a field $F$ are determined by its $x$ components
via a transformation $T$ depending only on the transition map $\varphi = y \circ x^{-1}$ between
the charts themselves. If we assume that $T$ is functionally differentiable in $[F]_x$ then
\be 
[\dg F]_y = \dg[F]_y = \dg T([F]_x)
	    = DT \cont \dg[F]_x,
\ee
where $DT$ is the derivative of $T$ and $\cont$ denotes contraction.
That is, $\dg F$ transforms according to the linearization of the transformation of $F$.
In the cases of interest to us $[F]_y$ is a function only of $[F]_x$ at the same manifold
point. That is, $[F]_y (y)= \tau([F]_x (\varphi^{-1} (y)))$ with $\varphi^{-1} (y)$ the $x$ coordinates
of the point defined by the values $y$ of the $y$ coordinates, and $\tau$ an ordinary
function of the space of components at a point to itself. Then the transformation law for
$\dg F$ reduces to
\be 
[\dg F]_y = D\tau \cont \dg[F]_x \circ \varphi^{-1},
\ee
with $D$ now the derivative in the space of components (which is finite dimensional for the
fields we encounter) and the contraction also taken in this space.

The variations $\dg^C F$ transform in precisely the same way. Let $x'$ and $y'$ be the
moving charts in $C$ formed by carrying $x$ and $y$ along the flow of $C$:
$x'_\lam \circ \Phi_{C\,\lam} = x$ and $y'_\lam \circ \Phi_{C\,\lam} = y$.
Then, by (\ref{delta_moving}), $[\dg^C F]_x = [\dg^C F]_{x'} = \dg[F]_{x'}$
and $[\dg^C F]_y = \dg[F]_{y'}$ at $\lam = 0$. But the transition map from
the $x'$ to the $y'$ chart is unaffected by the flow. It is just the transition
map $\varphi$ from $x$ to $y$ coordinates. The transformation from $x'$
to $y'$ components is thus also the same as that from $x$ to $y$ components:
$[F]_{y'} = T([F]_{x'})$. It follows that
$\dg^C F$ transforms according to
\be
[\dg^C F]_y = \dg[F]_{y'} = \dg T([F]_{x'})
	    = DT \cont \dg[F]_{x'} = DT \cont [\dg^C F]_{x},
\ee 
just like $\dg F$.


The variations of several integrals are evaluated in the present work. In particular, part of
the symplectic 2-form on $\cN_R$ is obtained by varying the bulk term in the symplectic
potential on $\cN_R$. The latter is an integral of the form $\int_{\cN_R} F$ where the integrand
is represented by the 3-form $F$. Using the $a_R$ chart this integral may be expressed as
$\int_{y[S_0]} \int_{r_0}^1 [F]_a dr d^2y$, an integral over a domain in $\mathbb{R}^3$.
Its variation is therefore
\be	\label{delta_int_a}
\dg \int_{\cN_R} F = \int_{y[S_0]} \int_{r_0}^1 \dg [F]_a dr d^2 y
-  \int_{y[S_0]} \dg [r_0]_y [F]_a|_{r = r_0} d^2 y.
\ee
Here $\dg [r_0]_y$ is the variation of the scalar $r_0$ at constant $y$.
This variation may be expressed in a coordinate independent way in terms of the
comoving variation $\dg^a$ associated with the $a_R$ chart:
\be 	\label{delta_int_a2}
\dg \int_{\cN_R} F = \int_{\cN_R} \dg^a F - \int_{S_0} \dg [r_0]_y \di_r \cont F,
\ee
with $S_0$ oriented so that $\int_{S_0} dy^1\wedge dy^2 = \int_{y[S_0]} d^2y$.

\end{document}